\documentclass{ametsocV6.1}
\usepackage{gensymb}
\usepackage{caption}
\usepackage{subcaption}
\usepackage{chngcntr}
\usepackage{tabularx}
\usepackage{xcolor}

\title{Tropical temperature distributions over a wide range of climates: theory and idealized simulations}

\authors{Joshua A. M. Duffield\aff{a} and Michael P. Byrne\aff{a}\correspondingauthor{Joshua A. M. Duffield, jamd1@st-andrews.ac.uk} 
}

\affiliation{\aff{a}{School of Earth and Environmental Sciences, University of St Andrews, St Andrews, UK}
}

\abstract{
Understanding future changes in temperature variability and extremes is an important scientific challenge with societal impacts. Here the responses of daily near-surface temperature distributions to climate warming is explored using an idealized GCM. Simulations of a wide range of climate states are performed using both a slab-ocean aquaplanet configuration and a simple continental configuration with a bucket-style model for land hydrology. In the tropics, the responses of extreme temperatures to climate change contrast strongly over land and ocean. Over land, warming is amplified for hot days relative to the average summer day. But over ocean, warming is suppressed for hot days, implying a narrowing of the temperature distribution.
Previous studies have developed theories based on convective coupling to interpret changes in extreme temperatures over land. Building on that work, the contrasting temperature distribution responses over land and ocean are investigated using a novel theoretical framework based on local convective coupling. The theory highlights five physical mechanisms with the potential to drive differential warming across temperature percentiles: free-tropospheric temperature change, relative humidity change, convective available potential energy (CAPE) change, the hot-get-hotter mechanism, and the drier-get-hotter mechanism. Hot days are relatively dry over land due to limited moisture availability, which drives the drier-get-hotter mechanism and amplified warming of the warm tail of the distribution. This mechanism is the primary factor explaining the contrasting responses of hot days over land and ocean to climate change. But other mechanisms also contribute to changing the temperature distribution, with changes in free-tropospheric temperature and surface relative humidity having large influences (which partially cancel).} 

\begin{document}

\maketitle

%
%
%
%
%

%
\section{Introduction}
\label{sect:intro}

\noindent Climate change affects not only mean temperatures but also has the potential to change the shape of temperature distributions. For example, climate models and observations show changes in summertime temperature variability in response to global warming \citep{fischer_future_2009,mckinnon_changing_2016}, including changes in higher-order moments like skewness \citep{tamarin-brodsky_changes_2020}. In general, changes in both the mean and shape of the temperature distribution affect extremes; for example, hot days may warm faster or slower than the average day depending on how the variance of the distribution evolves in a changing climate. Understanding how the temperature distribution -- and consequently the probability of extreme temperatures -- responds to climate change is an important scientific question with societal implications. On land, for example, hot extremes drive increased wildfire risk and this risk is projected to increase over the 21st century, especially in the northern hemisphere \citep{flannigan_global_2013}. Hot extremes also affect agriculture \citep{vogel_effects_2019}, health \citep{gasparrini_impact_2011}, and human infrastructure \citep{perkins_review_2015}. In ecology, changes in warm (and cold) extremes are crucial for understanding changes in species' performance with global warming \citep{vasseur_increased_2014}. And in the oceans, periods of extreme surface warmth (known as marine heatwaves) have numerous impacts on biodiversity, including mass coral bleaching \citep{hughes_global_2017}. Marine heatwaves have become longer and more frequent over the past century \citep{oliver_longer_2018}, and this trend is projected to accelerate as global warming intensifies \citep{frolicher_marine_2018}.

Physical understanding of how temperature responds to climate change is developing. Recent advances have narrowed uncertainty in the global-mean surface temperature response to radiative forcing \citep{sherwood_assessment_2020}, and key regional patterns of warming -- including polar amplification and the land-ocean warming contrast -- are well understood \citep[e.g.,][]{pithan_arctic_2014,byrne_trends_2018}. Beyond mean temperatures, progress is also being made in understanding how higher-order moments of the temperature distribution respond to climate warming. In mid-latitudes, on synoptic timescales, horizontal advection of heat is known to be a dominant process affecting temperature variability, particularly in winter \citep{holmes_robust_2016}. By linking advection to the reduction in meridional temperature gradient associated with polar amplification, it is expected that wintertime temperature variance will decrease as climate warms \citep{screen_arctic_2014,schneider_physics_2015}. Assuming a Gaussian temperature distribution this would lead to less frequent extremes in the future, but changes in skewness can alter this conclusion \citep{tamarin-brodsky_dynamical_2019}. Advection can also affect summertime temperature distributions, for example the rapid warming of Iberia and north Africa has been hypothesized to contribute to the amplified warming of hot days in north-west Europe \citep{patterson_north-west_2023}. More generally, advective arguments have been used to interpret how the skewness of mid-latitude temperature responds to climate change, both in summertime and wintertime \citep{tamarin-brodsky_changes_2020}.

Other processes, beyond advection, are important for understanding how mid-latitude hot extremes change with global warming \citep{linz_framework_2020}. Once an anticyclone is established, radiative processes are important for raising temperatures: subsidence inhibits cloud formation to bring clear skies and strong shortwave fluxes into the surface \citep{tian_radiation_2023}. Land-atmosphere feedbacks can also be important, with extreme temperatures affected by a well-established feedback loop \citep[e.g.,][]{fischer_future_2009,seneviratne_investigating_2010}: reduced soil moisture decreases evapotranspiration and latent cooling of the land surface, causing an increase in temperature. The raised temperature, in turn, leads to higher evaporative demand, drying soils further. This feedback loop can progress until the soil dries completely, at which point temperature increases cannot be damped further by increased evapotranspiration. The increased sensible heat flux also causes the atmospheric boundary layer to deepen, helping to maintain high temperatures overnight because of reduced entrainment of cold air from above \citep{miralles_mega-heatwave_2014}. Consistent with the mechanism described above, simulations with prescribed decreases in soil moisture show stronger warming of extreme temperatures relative to simulations with fixed soil moisture \citep{seneviratne_impact_2013}. The relative importance of the various processes driving extreme mid-latitude temperatures varies spatially \citep{rothlisberger_quantifying_2023}, with advection associated with anticyclones dominant over oceans. But over mid-latitude land, adiabatic warming associated with anticyclones and diabatic heating associated with shortwave radiation and turbulent surface fluxes are typically more influential than advection \citep{rothlisberger_quantifying_2023}.



In the tropics, where the effects of advection are relatively weak \citep{holmes_robust_2016} and where models project amplified warming of hot days \citep{mckinnon_et_al_2024}, an alternative paradigm is emerging to understand extreme temperatures in a changing climate. Together, frequent atmospheric convection \citep[convective quasi-equilibrium (CQE);][]{emanuel_large-scale_1994} and weak temperature gradients in the tropical free troposphere \citep[WTG;][]{sobel_modeling_2000} couple moist conserved quantities -- such as moist static energy (MSE) -- over land and ocean regions \citep{byrne_landocean_2013,zhang_how_2020}. This dynamical coupling implies that changes in near-surface MSE are approximately equal over land and ocean \citep{byrne_understanding_2016,berg_landatmosphere_2016,byrne_trends_2018,byrne_et_al_2024}. Based on this conceptual model, quantitative theories to understand the responses of tropical-mean temperature over land, extreme temperatures, and heat stress to climate change have been developed \citep{byrne_trends_2018,byrne_amplified_2021,zhang_projections_2021,duan2024moist}. Recently this paradigm -- which assumes that convection over land tightly couples the boundary layer to the free troposphere -- has been extended to mid-latitude regions, where it has been used to derive an upper bound for near-surface extreme temperatures as a function of free-tropospheric temperature on hot days \citep{zhang_upper_2023}.

Motivated by this work demonstrating a strong influence of convection on temperature distributions and by projections showing amplified warming of hot days in the tropics \citep{byrne_amplified_2021,mckinnon_et_al_2024}, here we explore in detail the sensitivity of tropical near-surface temperature distributions to climate change. We develop theory to predict how the temperature distribution responds to warming in the convective limit, and use the theory to interpret the behaviors of land and ocean tropical temperature distributions in idealized GCM simulations over a wide range of climates (cold to hot). This methodology has previously been applied to a variety of problems in climate dynamics including extreme precipitation \citep{ogorman_scaling_2009}, the Hadley circulation \citep{levine_response_2011}, mid-latitude temperature variability \citep{schneider_physics_2015}, and the land-ocean warming contrast \citep{byrne_landocean_2013}. We begin by describing the idealized GCM and simulations performed (section \ref{sect:model}) followed by a discussion of the variation in temperature distributions across these simulations (section \ref{sect:results_describe}). We then develop the convective theory (sections \ref{sect:theory} and \ref{sect:theory_explain}) and apply it to the simulations (sections \ref{sect:results_explain} and \ref{sect:wtg}). We conclude with a summary and discussion (section \ref{sect:summary}).

\section{Simulations}
\label{sect:model}

\noindent We perform idealized simulations using the Isca climate model \citep{vallis_isca_2018}. Simulations are run at T42 horizontal spectral resolution (approximately $2.8^\circ \times 2.8^\circ$) with 25 vertical $\sigma$-levels and a simplified Betts-Miller convection scheme following \citet{frierson_dynamics_2007}. All simulations are spun up for a period of 720 days, with data from the subsequent 1800 days used in the analyses described below.

Top-of-atmosphere (TOA) insolation, $S_{TOA}^{\downarrow}$, is prescribed as a function of latitude with both a seasonal and diurnal cycle. The solar constant and obliquity are set to $1368.22 \, \rm{W/m^2}$ and 23.439$^\circ$, respectively. Radiative transfer is modeled using a semi-gray scheme with prescribed longwave and shortwave optical thicknesses \citep{frierson_gray-radiation_2006}. Note that, in this simple scheme, interactions between radiation, clouds, and water vapor are absent. 

The longwave and shortwave optical thicknesses are specified similarly to \citet{ogorman_hydrological_2008} but with different parameters. The longwave optical thickness is specified by $\tau=\kappa \tau_{ref}$, with the optical thickness of a reference simulation, $\tau_{ref}$, given by:

\begin{equation}
\tau_{ref}(p,\phi)=\left[f_l(p/p_0) + (1-f_l)(p/p_0)^4\right]\left[\tau_e + (\tau_p-\tau_e)\sin^2\phi\right],
\label{eqn:optical_depth_lw}
\end{equation}
where $f_l=0.1$, $p$ is pressure, $p_0=1000$ hPa is a reference pressure, $\phi$ is latitude, and the longwave optical thicknesses at the equator and poles are $\tau_e=6.0$ and $\tau_p=1.5$, respectively. A wide range of climates is simulated by scaling the reference longwave optical thickness via the parameter $\kappa$. Simulations with eight $\kappa$ values are performed ($\kappa=0.6, 0.8, 1.0, 1.5, 2.0, 2.5, 3.0$ and $3.5$).

Absorption of solar radiation in the atmosphere is imposed such that the downward shortwave flux at a given pressure level and latitude is:

\begin{equation}
S^{\downarrow}(p)=S^{\downarrow}_{TOA}\exp\left[-\tau_s(p/p_0)^4\right],
\label{eqn:optical_depth_sw}
\end{equation}
where $\tau_s=0.2$ and $S_{TOA}^{\downarrow}$ is the incident shortwave flux at TOA. Shortwave radiation is only absorbed from the downward stream and reflection occurs only at the surface, such that the upward shortwave flux at TOA is $S_{TOA}^{\uparrow} = \mathcal{A} S_{TOA}^{\downarrow} \exp[-\tau_s]$, where $\mathcal{A} = 0.31$ is the prescribed, spatially-uniform surface albedo. 

Surface turbulent fluxes are computed using bulk aerodynamic formulae together with Monin-Obukhov similarity theory.

\subsection{Aquaplanet configuration}

Aquaplanet simulations are performed using a saturated surface with a heat capacity equal to a 1 m layer of liquid water. There are no prescribed horizontal heat transports in this surface layer. A small surface water depth is used to ensure convection is triggered for a significant fraction of the year. The reference simulation ($\kappa = 1.0$) has a global-mean near-surface temperature similar to that of present-day Earth (284 K), whilst the $\kappa=0.6$ and $\kappa=3.5$ simulations have global temperatures of 276 K and 302 K, respectively. 



\subsection{Land configuration}

Simulations are also performed with a meridional land band of width $60\degree$ in longitude and spanning all latitudes \citep[a setup similar to that used by][]{byrne_landocean_2013}. In these simulations, the only difference between land and ocean is how the surface evaporative flux is parameterized. The land simulations use an interactive bucket-style model for hydrology including a vegetation prefactor, $C_V=0.1$, following \citet{pietschnig_response_2021}. The effect of the vegetation prefactor is to reduce the surface relative humidity increase with warming for the coldest summer days, inline with expectations from CMIP6 simulations [see Extended Data Fig.2 in \citet{byrne_amplified_2021}]. In this formulation the bucket depth at each gridpoint (denoted $W$), which is analogous to soil moisture, varies in response to the local balance of precipitation and evaporation. Land evaporation, $E_L$, depends on bucket depth and is given by $E_L = C_V C_L(W) E_0$, where $C_L(W)$ is the surface moisture conductivity and $E_0$ is the potential evaporation rate (i.e., the evaporation that would occur from from a saturated surface). Following \citet{manabe1969climate}, we specify $C_L = 1$ for $W \geq 0.75 W_f$ and $C_L = W/(0.75W_f)$ for $W < 0.75 W_f$, where $W_f$ is the field capacity. Land simulations are performed for the same range of longwave optical thickness parameters, $\kappa$, as for the aquaplanet simulations.


\subsection{Notes on the analyses}

Daily-mean quantities are used throughout the article and $\chi[x]$ refers to a variable $\chi$ conditioned on the $x$th percentile of near-surface temperature. Specifically, $\chi[x]$ is computed by taking the average of $\chi$ across all days and gridpoints with near-surface temperature in the percentile range between $x-0.5<x\leq x+0.5$. The $x^{th}$ percentile of $\chi$ is denoted $\chi(x)$ and in general, $\chi(x) \neq \chi[x]$ except for when $\chi$ is near-surface temperature, $T_s$. We aggregate over longitude and time prior to computing percentiles and mean quantities (with the latter denoted $\overline{\chi}$). When averaging over a range of latitudes, we first compute percentiles or mean quantities at each latitude individually before averaging meridionally with area weighting. The results focus on two vertical levels: (i) the near surface (hereafter referred to as ``surface'' and denoted by a subscript $s$), taken to be the lowest vertical level ($\sigma=0.995$); and (ii) the free troposphere (denoted by a subscript $FT$), taken to be the $\sigma=0.487$ level. For the land simulations, only longitudes corresponding to land are included in the calculation of percentiles and mean quantities. In subsequent analyses we aggregate data over a six month summer (MJJASO for the northern hemisphere; NDJFMA for the southern hemisphere) and consider only tropical latitudes ($20\degree$S to $20\degree$N) where convection is expected to be active.

Throughout this study, figures show results combined from cold ($\kappa=0.6, 0.8, 1.0$ and $1.5$) and hot simulations ($\kappa=1.5, 2.0, 2.5, 3.0$ and $3.5$). This is done using linear regression; for example, to plot $\delta \chi_1[x]/\delta \overline{\chi}_2$ versus $x$ for given variables $\chi_1$, $\chi_2$, where $\delta$ means change with warming, we perform a least squares regression of $\chi_1[x]$ versus $\overline{\chi}_2$ at each percentile, $x$. For both the cold and hot simulations, the regression is forced to go through the $\kappa=1.5$ value of $\left(\overline{\chi}_2, \chi_1[x]\right)$.



\section{Summertime tropical temperature distributions in idealized simulations}
\label{sect:results_describe}

\begin{figure}
\center
\includegraphics[width=35pc]{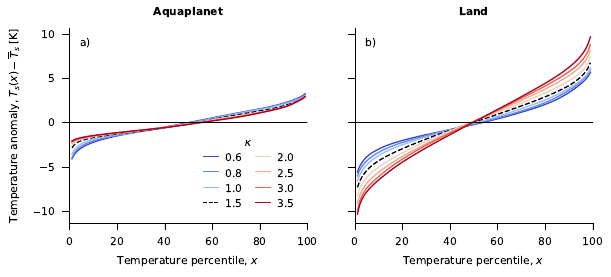}
\caption{Anomalies of percentiles of surface temperature, $T_s(x)$, relative to the time- and zonal mean temperature, $\overline{T}_s$, for each (a) aquaplanet and (b) land simulation, as a function of surface temperature percentile, $x$. For the land simulations, only land coordinates are considered. A widening of the land temperature distribution with climate warming is evident -- with high percentiles warming, and low percentiles cooling, relative to the median temperature -- as is a narrowing of the ocean distribution. In this and subsequent figures, all quantities are spatially averaged from $20\degree$S to $20\degree$N and temporally averaged over the summer (6 warmest months).}\label{fig:temp_surf_anom}
\end{figure}

\noindent In the aquaplanet simulations there is a weak narrowing with warming of the summertime tropical temperature distribution, with warming of low percentiles and cooling of high percentiles relative to the mean temperature (Fig. \ref{fig:temp_surf_anom}a). Narrowing of the aquaplanet distribution with warming is described by the standard deviations across all percentiles, which decreases from $1.66$ K to $1.20$ K between the $\kappa=0.6$ and $3.5$ simulations. Narrowing of the distribution is not symmetric either side of the mean temperature, with greater narrowing on the cool side compared to the warm side -- the standard deviation decreases by $0.50$ K between $\kappa=0.6$ and $\kappa=3.5$ for percentiles $x\leq50$, but only by $0.03$ K for $x \geq 50$.


In the land simulations, the summertime tropical temperature distribution is broader than the aquaplanet distribution for any given $\kappa$ simulation (Fig. \ref{fig:temp_surf_anom}b). Heat capacities and albedos in the land and aquaplanet simulations are identical, implying the broader land temperature distribution is caused by limited evaporation and surface dryness. In contrast to the aquaplanet, the land temperature distribution robustly widens with warming, and this widening accelerates as the climate is warmed: the standard deviations across all percentiles being $2.38$ K, $3.15$ K, and $4.65$ K for the $\kappa=0.6$, $1.5$, and $3.5$ simulations, respectively. The widening of the land distribution is more symmetric compared to the aquaplanet simulations, but is larger on the cool side -- the standard deviation increases by $1.37$ K between the $\kappa=0.6$ and $\kappa=3.5$ runs for $x\leq50$, and by $0.91$K for $x \geq 50$. 


The responses to warming of the temperature distributions are quantified by a scaling factor, defined as the change in surface temperature at each percentile, $\delta T_s(x)$, normalized by the mean temperature change, $\delta \overline{T}_s$ (Fig. \ref{fig:sf_theory}). The scaling factor is computed for a range of simulations by regressing $T_s(x)$ versus $\overline{T}_s$ for each percentile $x$. Note this scaling factor is defined differently to that used in \citet{byrne_amplified_2021}. The narrowing (widening) of the distribution for the aquaplanet (land) simulations is apparent from $\delta T_s(x)/\delta \overline{T}_s >1$ for small (large) values of $x$. The more substantial changes in the shape of the land temperature distribution compared to the aquaplanet distribution are evident from the larger magnitude of $\delta T_s(x)/\delta \overline{T}_s -1$ for land in Figure \ref{fig:sf_theory}.

\section{Theory}
\label{sect:theory}

\noindent To gain physical insight into the response of tropical surface temperature distribution to climate change, we build on recent theoretical arguments [in particular \citet{byrne_amplified_2021}] to develop an extended theory to interpret the results from the idealized simulations described above. 

The theory of \citet{byrne_amplified_2021} is based on constraints imposed by atmospheric convection and weak temperature gradients on MSE and other moist adiabatically-conserved quantities [e.g., equivalent potential temperature \citep{byrne_ogorman_2013b} or wet-bulb temperature \citep{zhang_projections_2021}]. The idea is that frequent convection tightly connects the tropical boundary layer to the free troposphere, maintaining a lapse rate that is close to moist adiabatic \citep{emanuel_large-scale_1994}. This strong vertical coupling leads to co-variation of boundary-layer MSE with free-tropospheric saturation MSE \citep{nie2010observational,boos2015review}. Together with weak free-tropospheric temperature gradients, this vertical coupling implies that -- where convection is active -- boundary-layer anomalies in MSE are roughly equal across the tropics. 

Here we develop a related but distinct version of this theory, where we relax the assumption of weak free-tropospheric gradients and focus on the local vertical coupling between the boundary layer and free troposphere assuming frequent convection. In scenarios where convection is the dominant influence on lapse rates, the vertical temperature profile is close to moist adiabatic and the limit of ``strict'' convective equilibrium \citep[SCE;][]{emanuel2007quasi}. Going forward, we use SCE to refer to the scenario of a neutrally-stable moist adiabatic temperature profile [i.e., with a dry adiabatic lapse rate between the surface and lifting condensation level (LCL) and a saturated moist adiabatic lapse rate above the LCL]. The effects of other processes -- for example advection, diabatic heating, and soil moisture -- on temperature distributions and extremes have been studied extensively \citep[e.g.,][]{fischer2007soil,tamarin-brodsky_dynamical_2019,zeppetello2022physics,rothlisberger2023quantifying}. Relatively few studies have focused on the role of convection, though this has been an area of growing interest over recent years \citep{byrne_amplified_2021,zhang_upper_2023,duan2024moist,li2025atmospheric}. The influence of convection on temperature distributions is expected to be most relevant in the tropics but is also likely to be important over mid-latitude land in summertime, when the thermal stratification is strongly influenced by moist convection \citep{korty_climatology_2007}.



A straightforward way to quantify vertical coupling by convection is in terms of MSE, with near-surface MSE, $h_s$, related to saturation MSE in the free troposphere by:


\begin{equation}
h_s = h_{FT}^* + \epsilon,
\label{eqn:mse_coupling}
\end{equation}
where $h_s = c_p T_s + L_v q_s + gz_s$, $h_{FT}^* = c_p T_{FT} + L_v q_{FT}^* + gz_{FT}$, stars denote saturation values, and the other symbols have their usual meanings. The parameter $\epsilon$ represents a departure from the SCE limit, where $h_s=h_{FT}^*$ by definition. The specific heat capacity of air at constant pressure is given by $c_p = 1004.6$ J kg$^{-1}$ K$^{-1}$ and the latent heat of vaporization is given by $L_v = 2.5 \times 10^6$ J kg$^{-1}$. The departure from SCE, $\epsilon$, can be interpreted as a simple proxy for convective available potential energy (CAPE) and convective inhibition (CIN).


Equation (\ref{eqn:mse_coupling}) is our starting point for deriving a theory for the response of the surface temperature distribution to climate change. This expression involves two related free-tropospheric quantities, temperature and geopotential height, and when considering the response to climate change it is helpful for interpretation to combine the two terms. To relate free-tropospheric temperature to geopotential height, we assume hydrostatic balance and a constant lapse rate between the surface and free troposphere to obtain (see Appendix \ref{appendix:derivation} for details):

\begin{equation} \label{eqn:z_temp_ft_relation}
z_{FT} - z_s \approx \frac{R^{\dagger}}{g}(T_s + T_{FT}),
\end{equation}
where $R^{\dagger} = R\ln \left(p_s / p_{FT} \right)/2$ is a modified gas constant ($R=287.04$ J kg$^{-1}$ K$^{-1}$ is the regular gas constant for dry air). Substituting (\ref{eqn:z_temp_ft_relation}) into (\ref{eqn:mse_coupling}) and moving all surface quantities to one side, we derive a new expression for vertical coupling containing free-tropospheric variables that are only a function of free-tropospheric temperature:

\begin{equation}
(c_p - R^{\dagger})T_s + L_v q_s - \epsilon \approx (c_p + R^{\dagger}) T_{FT} + L_v q_{FT}^*,
\label{eqn:mod_mse_coupling}
\end{equation}
Taking changes in (\ref{eqn:mod_mse_coupling}) between climate states (i.e., between simulations) and introducing a ``pseudo'' relative humidity $r_s \equiv q_s / q_s^*$ following \citet{byrne_understanding_2016} (hereafter referred to as ``relative humidity''), we obtain:

\begin{equation} \label{eqn:mse_mod_change}
(c_p - R^{\dagger})\delta T_s + L_v q_s^*\delta r_s + L_v (r_s+\delta r_s) \delta q_s^* - \delta \epsilon \approx (c_p + R^{\dagger}) \delta T_{FT} + L_v \delta q_{FT}^*.
\end{equation}
Rearranging (\ref{eqn:mse_mod_change}), it is straightforward to obtain an expression for $\delta T_s$ in terms of the climatological state (i.e., $T_s$, $r_s$, and $T_{FT}$) and changes in relative humidity and free-tropospheric temperature. Doing this for both the time- and zonal-mean, conditioning on the $x$ percentile of surface temperature, and neglecting some small terms (see Appendix \ref{appendix:derivation}), we obtain an expression for the scaling factor:

\begin{equation} \label{eqn:sf_theory}
\frac{\delta T_s(x)}{\delta\overline{T}_s} \approx
\gamma_{\delta T_{FT}} \frac{\delta T_{FT}[x]}{\delta \overline{T}_s} 
+ \gamma_{\Delta T_s}\frac{\Delta T_s(x)}{\overline{T}_s}
- \gamma_{\delta r} \frac{\overline{T}_s}{\overline{r}_s} \frac{\delta r_s[x]}{\delta \overline{T}_s}
- \gamma_{\Delta r} \frac{\Delta r_s[x]}{\overline{r}_s}
+ \gamma_{\delta T_{FT}}\frac{\delta sCAPE[x]}{R^{\dagger} \delta \overline{T}_s},
\end{equation}
where $\Delta$ refers to the anomaly of a given variable $\chi$ relative to the mean, i.e. $\Delta \chi[x] = \chi[x] - \overline{\chi}$. Each term on the RHS of (\ref{eqn:sf_theory}) corresponds to a separate physical mechanism with the potential for driving differential changes in temperature across the distribution (i.e., across $x$). The strength of each mechanism depends on a dimensionless sensitivity parameter $\gamma$; these parameters depend only on mean surface temperature and relative humidity in the control climate and physical constants. The sensitivity parameters are almost always positive, though $\gamma_{\Delta T_s}$ and $\gamma_{\Delta r}$ can be negative under certain conditions. See Appendix \ref{appendix:derivation}, specifically equations (\ref{eqn:gamma_ft})--(\ref{eqn:gamma_r_anom}) for the mathematical expressions for the sensitivity parameters as well as Figure \ref{fig:sensitivity} for graphical representations.

In going from (\ref{eqn:mse_mod_change}) to (\ref{eqn:sf_theory}), we replaced $\epsilon$ with a simple CAPE-like quantity based on a single pressure level (see Appendix \ref{appendix:cape}):

\begin{equation} \label{eqn:cape}
sCAPE=R^{\dagger}(T_{FT,SCE} - T_{FT}),
\end{equation}
where $sCAPE$ denotes the simple CAPE-like quantity and $T_{FT,SCE}$ is the temperature at $p_{FT}$ following a moist adiabat from the surface. In general this temperature differs from the environmental $T_{FT}$ when $\epsilon \neq 0$; in the SCE limit $T_{FT}=T_{FT,SCE}$ and $sCAPE=0$.

\begin{figure}
\center
\includegraphics[width=35pc]{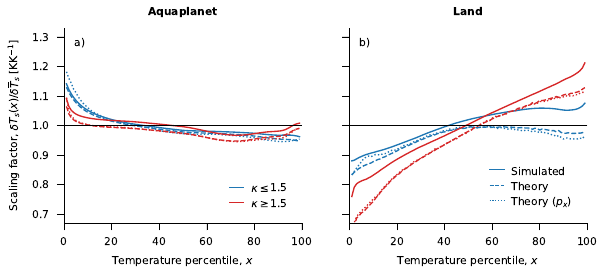}
\caption{Scaling factor for (a) aquaplanet and (b) land simulations as a function of surface temperature percentile. Scaling factors are shown for the four coldest (blue) and five hottest (red) simulations. Theory estimates of the scaling factors using (\ref{eqn:sf_theory}) and (\ref{eqn:sf_theory_p_x}) are shown by the dashed and dotted lines, respectively. The scaling factor is defined as the temperature at each percentile, $T_s(x)$, regressed against
 the mean temperature, $\overline{T}_s$, across the ranges of simulations shown on the legend of panel (a).}\label{fig:sf_theory}
\end{figure}

For the aquaplanet simulations, the theory captures very well the variation of the scaling factor with $x$ (see dashed lines in Fig. \ref{fig:sf_theory}a). The theory systematically underestimates the aquaplanet scaling factor for $\kappa \geq 1.5$, as is also seen for the land simulations (Fig. \ref{fig:sf_theory}b). The theory captures the key features of the land response: the hottest days warm more than the coldest days leading to a broadening of the distribution and the broadening is greater for $\kappa \geq 1.5$ (cf. red and blue lines in Fig. \ref{fig:sf_theory}b). However, the theory underestimates warming of the hottest days relative to the mean day; for $\kappa \geq 1.5$, it also underestimates the warming of the coldest days. A discussion of the approximations that lead to these discrepancies is provided in Appendix \ref{appendix:derivation}.

\begin{figure}
\center
\includegraphics[width=35pc]{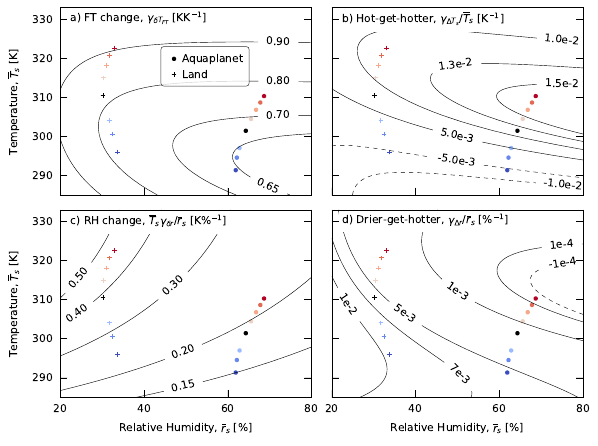}
\caption{Dependence of the sensitivity parameters, multiplied by the factor with which they appear in (\ref{eqn:sf_theory}), on mean surface temperature, $\overline{T_s}$ and relative humidity, $\overline{r_s}$.  I.e. these are the theoretical scaling factor contributions which result from each of the following having a value of 1: (a) rate of free-tropospheric temperature change, $\delta T_{FT}/\delta \overline{T}_s$ [KK$^{-1}$] , (b) surface temperature anomaly, $\Delta T_s$ [K], (c) rate of relative-humidity change, $\delta r_s/\delta \overline{T}_s$ [$\%\text{K}^{-1}$], and (d) relative humidity anomaly, $\Delta r_s$ [$\%$]. The dashed contours in (b) and (c) indicate negative values. The summer values averaged over the tropics for the eight aquaplanet (land) simulations are shown by dots (crosses) with the same colors as in Fig. \ref{fig:temp_surf_anom}. Equations for each $\gamma$ are given in Appendix \ref{appendix:derivation} (\ref{eqn:gamma_ft}-\ref{eqn:gamma_r_anom}).}\label{fig:sensitivity}
\end{figure}

Next we discuss the five terms on the RHS of (\ref{eqn:sf_theory}) and the associated physical mechanisms leading to differential warming across the temperature distribution.

\section{Physical mechanisms for differential warming}
\label{sect:theory_explain}

\begin{figure}
\center
\includegraphics[width=35pc]{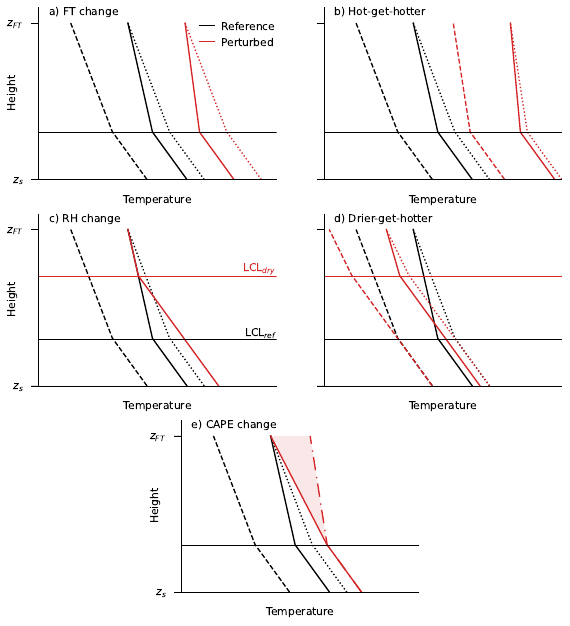}
\caption{Schematic diagrams illustrating the five mechanisms which can drive differential warming across temperature percentiles: (a) free-tropospheric temperature change; (b) hot-get-hotter; (c) relative humidity change; (d) drier-get-hotter; and (e) CAPE change. Dashed and solid lines show temperature profiles in cold and warm climates, respectively, while dotted lines show the warm-climate profiles neglecting changes in lapse rate. Black lines (same in each panel) show a reference scenario with moist adiabatic lapse rates in each climate and a prescribed free-tropospheric warming, but all other terms on the RHS of (\ref{eqn:sf_theory}) set to zero. Red lines show the perturbed scenarios in which each of the five terms on the RHS of (\ref{eqn:sf_theory}) is independently perturbed relative to the reference scenario. In panel (e), the temperature profile in the warm climate (red solid line) differs from the moist adiabatic profile (red dashed-dotted line), with associated CAPE (indicated by the shaded red area). Note these schematics are designed to aid understanding and are not quantitatively accurate.}\label{fig:schematic}
\end{figure}

Each term on the RHS of (\ref{eqn:sf_theory}) describes a physical mechanism capable of driving differential warming across percentiles of the temperature distribution. These mechanisms can be interpreted through consideration of lapse rates and their responses to warming, as shown in Figure \ref{fig:schematic}. In each schematic, warming under two scenarios is compared: (i) a reference scenario in which the free troposphere is prescribed to warm by a given magnitude ($\delta T_{FT}=\delta T_{FT,ref} > 0$) but all other terms on the RHS of (\ref{eqn:sf_theory}) are set to zero; and (ii) a perturbed scenario in which each of the five terms on the RHS of (\ref{eqn:sf_theory}) are perturbed relative to the reference scenario so as to illustrate how each mechanism can drive differential warming. Each mechanism is described in detail below.

\begin{figure}
\center
\includegraphics[width=19pc]{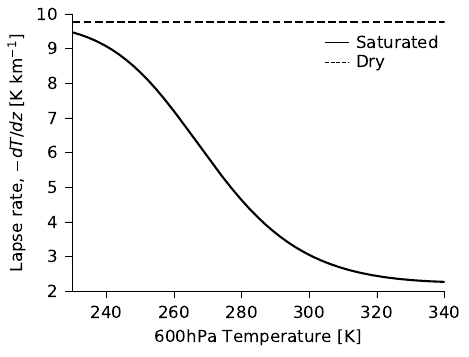}
\caption{Dependence of the saturated moist (solid) and dry (dashed) adiabatic lapse rates on temperature at the 600 hPa level. The saturated moist adiabatic lapse rate is computed using Eq. (9.42) from \citet{holton_introduction_2004}.}\label{fig:lapse_rate}
\end{figure}

\subsection{Free-tropospheric temperature change mechanism}

This mechanism is straightforward: in the limit of constant lapse rates, percentiles that warm more in the free troposphere also warm more at the surface (cf. black and red dotted lines in Fig. \ref{fig:schematic}a). However, lapse rates generally change with warming: the larger the warming, the larger the decrease in the saturated portion (above the LCL) of the moist adiabatic lapse rate (Fig. \ref{fig:lapse_rate}). This means surface warming will be smaller than free-tropospheric warming, all else constant, except in the dry limit where warming is equal at all vertical levels (because the dry adiabatic lapse rate is independent of temperature; see Fig. \ref{fig:lapse_rate}). 



Differential warming in the free troposphere across temperature percentiles can clearly lead to differential surface warming. But the strength of this mechanism depends on changes in the moist adiabat, which depend on temperature (Fig. \ref{fig:lapse_rate}) and surface relative humidity. This sensitivity is quantified through the $\gamma_{\delta T_{FT}}$ parameter [see (\ref{eqn:sf_theory})], which is positive and generally larger in dry and hot climates (Fig. \ref{fig:sensitivity}a). This is because the lapse rate is less sensitive to warming in dry, hot climates; consequently, surface warming is similar to free-tropospheric warming. The temperature dependence of $\gamma_{\delta T_{FT}}$ is clear from Figure \ref{fig:lapse_rate}; the saturated lapse rate decreases less for a given warming at high temperatures. The dependence of $\gamma_{\delta T_{FT}}$ on relative humidity is also intuitive: lower relative humidity implies a higher LCL and deeper boundary layer where temperature follows a dry adiabatic lapse rate. The dry adiabat is constant with warming leading to a larger sensitivity of surface temperature to a change in free-tropospheric temperature (i.e., larger $\gamma_{\delta T_{FT}}$). 


\subsection{Hot-get-hotter mechanism}

The ``hot-get-hotter'' mechanism, quantified by $\gamma_{\Delta T_s} \Delta T_s(x)/\overline{T}_s$ [see (\ref{eqn:sf_theory})], implies that surface temperature percentiles which are climatologically hotter than average [i.e., $\Delta T_s(x) > 0$] experience amplified warming relative to the average day in a changing climate. 

The hot-get-hotter mechanism is illustrated in Figure \ref{fig:schematic}b. The climatologically hotter percentiles warms more than the reference scenario at the surface, despite the same free-tropospheric warming, due to weaker decreases in lapse rate with warming. This is apparent from the temperature profile in the warmer climate (solid line) being closer to the fixed lapse rate profile (dotted line) for the climatologically hotter day (red) compared to the reference scenario (black) representing, for example, the average temperature profiles. Similar to the lapse rate contribution to the free tropospheric temperature change mechanism, the hot-get-hotter mechanism arises because the sensitivity to warming of the saturated moist adiabatic lapse rate is larger for cooler days (Fig. \ref{fig:lapse_rate}), implying that, for a given warming, the lapse rate decreases less for hotter days. 

The strength of this mechanism is quantified by the parameter $\gamma_{\Delta T_s}$, which becomes negative (i.e., ``colder-get-hotter'') for low surface temperature (Fig. \ref{fig:sensitivity}b). This change in sign occurs because for cold climates ($T\lesssim270$ K), the gradient of the saturated moist adiabatic lapse rate increases with temperature (Fig. \ref{fig:lapse_rate}), such that for a given warming, the lapse rate decreases more for hotter days. 

In the hot-get-hotter regime, the sensitivity increases in hotter and more humid climates (except for very hot climates with $\overline{T}_s\gtrsim320$ K). More humid climates are more sensitive because this mechanism depends entirely on how the moist adiabatic lapse rate changes with warming, and in more humid climates the LCL is lower so a larger proportion of the temperature profile follows the saturated lapse rate. The temperature dependence is more complicated: in general, a larger sensitivity arises from a larger second derivative of the saturated lapse rate with respect to temperature as this increases the difference in lapse rate changes for a given warming between two percentiles with different starting temperatures.

\subsection{Relative humidity-change mechanism}

The relative humidity-change mechanism, $-\gamma_{\delta r}\frac{\overline{T}_s}{\overline{r}_s}\delta r_s[x]/\delta \overline{T}_s$ [see (\ref{eqn:sf_theory})], describes how drying ($\delta r_s<0$) leads to amplified surface warming compared to the reference scenario of constant relative humidity.

This mechanism is illustrated in Figure \ref{fig:schematic}c. The decrease in surface relative humidity with warming causes the LCL to rise (cf. solid red and black lines). This deepening of the dry boundary layer results in more of the temperature profile following a dry adiabat rather than a saturated moist adiabat, implying greater surface warming compared to the reference scenario.

The strength of this mechanism, quantified by the $\gamma_{\delta r}$ parameter, increases in drier and hotter climates (Fig. \ref{fig:sensitivity}c). Amplified warming from this mechanism arises from changing a section of the vertical temperature profile from a saturated  moist adiabatic lapse rate to a dry adiabatic lapse rate. The amplified warming is thus accentuated by either increasing the depth of this section or increasing the difference between the moist and dry lapse rates. This lapse-rate difference increases with temperature (Fig. \ref{fig:lapse_rate}), explaining the larger sensitivity parameter in warmer climates (Fig. \ref{fig:sensitivity}c). The increased sensitivity in drier climates is explained by the LCL being more sensitive to relative humidity decreases in drier climates [see Fig. 3a in \citet{romps_exact_2017}]. Hence, for a given relative humidity decrease, the LCL rises more in a drier climate and a greater section of the temperature profile switches to a dry adiabat.

\subsection{Drier-get-hotter mechanism}

The ``drier-get-hotter'' mechanism, $-\gamma_{\Delta r} \Delta r_s[x]/\overline{r}_s$ [see (\ref{eqn:sf_theory})], implies that climatologically dry percentiles of the temperature distribution (defined as percentiles with $\Delta r_s[x] <0$), will experience amplified warming relative to the mean, all else equal. A similar mechanism was described by \citet{byrne_amplified_2021}; here we derive a modified version of this drier-get-hotter mechanism in the context of local convective coupling.

The drier-get-hotter mechanism is illustrated in Figure \ref{fig:schematic}d. The perturbed scenario (red lines) has lower surface relative humidity compared to the reference scenario, causing the LCL in the perturbed scenario to be higher. There is a height range between the LCLs where the temperature profile in the perturbed scenario follows a dry adiabat while the reference profile follows a saturated moist adiabat. With warming the saturated moist adiabatic lapse rate decreases but the dry adiabatic lapse rate remains constant (Fig. \ref{fig:lapse_rate}), implying larger surface warming for the perturbed (dry) scenario as it has a deeper dry boundary layer.

The strength of this mechanism, quantified by the $\gamma_{\Delta r}$ parameter, increases in drier and colder climates (Fig. \ref{fig:sensitivity}d). Differential surface warming across temperature percentiles arises from surface dryness inducing a difference in LCL heights across percentiles. The magnitude of this differential warming is therefore increased by either increasing the LCL height difference or increasing the magnitude of the decrease in the saturated moist adiabatic lapse rate with warming. The explanation for the strength of this mechanism in dry climates is the same as for the relative humidity-change mechanism: for a given relative humidity anomaly, $\Delta r_s$, the LCL anomaly is larger in a dry climate. The explanation for the strength of the mechanism in cold climates is similar to the hot-get-hotter mechanism: for a given warming, the saturated moist adiabatic lapse rate decrease is larger in a colder base climate (Fig. \ref{fig:lapse_rate}), which increases the contrast in the lapse rate changes between the perturbed and reference scenarios in the vertical layer between the LCLs.

\subsection{CAPE-change mechanism}

The CAPE-change mechanism, $\gamma_{\delta T_{FT}}\delta sCAPE[x]/(R^{\dagger}\delta \overline{T}_s)$ [see (\ref{eqn:sf_theory})], describes how an increase in simple CAPE with warming ($\delta sCAPE/\delta \overline{T}_s>0$) leads to amplified surface warming compared to the reference scenario with zero CAPE change. This mechanism is shown in Figure \ref{fig:schematic}e, where the perturbed scenario has $\delta sCAPE>0$ (red lines). In contrast to the previous mechanisms, a CAPE change implies that the response of the lapse rate to warming deviates from that of a moist adiabat. In the schematic (Fig. \ref{fig:schematic}e), this is illustrated by the lapse rate in the perturbed scenario decreasing by less than the moist adiabatic lapse rate above the LCL (cf. red and black solid lines). Below the LCL, the temperature profile in the perturbed scenario follows a dry adiabat. The red dashed-dotted line indicates the moist adiabatic profile in the warm climate given the surface temperature and relative humidity. The CAPE is related to the temperature difference between the perturbed and moist adiabatic profiles (shaded area in Fig. \ref{fig:schematic}e), and the larger this difference the larger the amplified surface warming compared to the reference scenario with no CAPE change. Physically, one explanation for why tropical temperature profiles deviate from moist adiabats is entrainment \citep[e.g.,][]{singh_influence_2013}: if dry air is entrained into a convective plume, the lapse rate tends more towards the dry adiabat than the saturated moist adiabat.

The strength of this mechanism is controlled by the same sensitivity parameter, $\gamma_{\delta T_{FT}}$, as the free-tropospheric temperature change mechanism, which is larger in dry and cold climates (Fig. \ref{fig:sensitivity}a). In our formulation, a positive CAPE change is equivalent to a further increase in free-tropospheric temperature, while maintaining a moist adiabatic profile (as far as the surface temperature is concerned, the red solid and dashed-dotted lines in Fig. \ref{fig:schematic}e are equivalent). The transmission of this additional change in free-tropospheric temperature through the change in moist adiabatic profile (black solid to red dot-dashed line in Fig. \ref{fig:schematic}e) to the surface is  then quantified through $\gamma_{\delta T_{FT}}$.

In the next section, the theoretical mechanisms described above will be applied to interpret the responses of tropical summer temperature distributions to climate change across the idealized simulations.

\section{Theory applied to idealized simulations}
\label{sect:results_explain}


In this section, each term in the theoretical scaling factor (\ref{eqn:sf_theory}) is computed for the idealized simulations and its importance for driving differential warming across percentiles is assessed.


\begin{figure}
\center
\includegraphics[width=35pc]{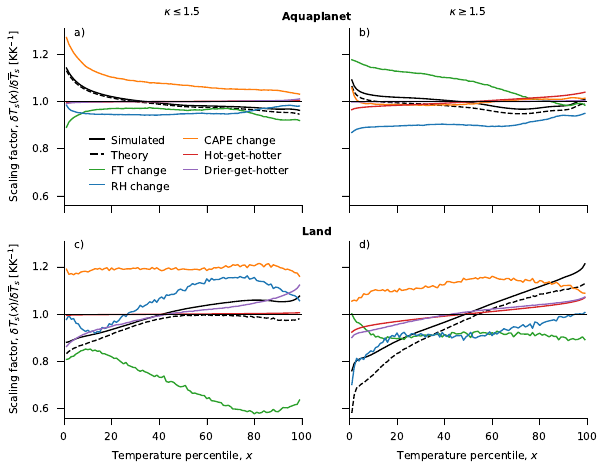}
\caption{Contribution of each of the five terms in the theory (\ref{eqn:sf_theory}) to the scaling factors versus temperature percentile. The sum of the five terms corresponds to the full theory (black dashed line), which deviates from the simulated scaling factors (black solid line). The breakdown for the four coldest (a, c) and five warmest (b, d) aquaplanet and land simulations are shown. Note that 1 has been added to all contributions except FT change (an FT change contribution of 1 across all percentiles corresponds to surface warming but no differential surface warming; while for any other mechanism, this would correspond to no surface warming at all.)}\label{fig:sf_theory_breakdown}
\end{figure}

\begin{figure}
\center
\includegraphics[width=35pc]{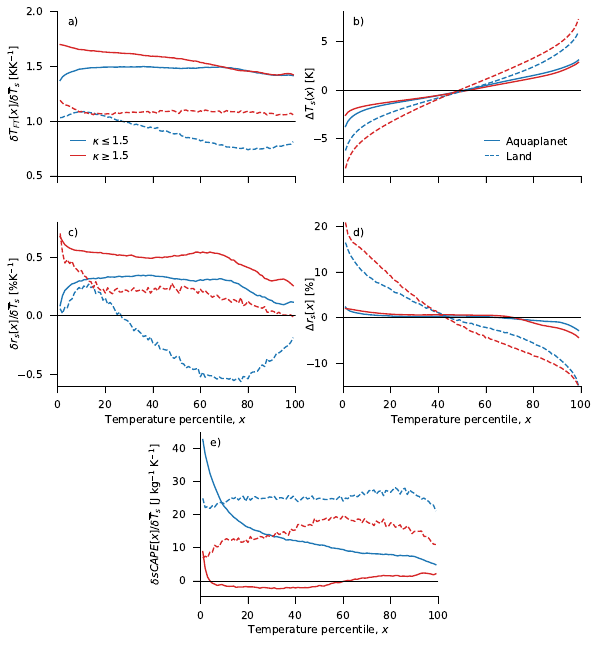}
\caption{Various physical quantities that contribute to the scaling factor (\ref{eqn:sf_theory}), each conditioned on -- and plotted versus -- temperature percentile: (a) free-tropospheric temperature change normalized by mean surface temperature change; (b) surface temperature anomaly; (c) normalized relative humidity change; (d) relative humidity anomaly; and (e) normalized $sCAPE$ change. Solid lines show data for the aquaplanet simulations and dashed lines show the land simulations. In each case, regression is performed to show a single line for the four coldest (blue) and five hottest (red) simulations. To obtain (b) and (d), the regression that is performed is $\overline{T}_s\Delta \chi[x]$ against $\overline{T}_s$ for $\chi \in \{T_s, r_s\}$.}\label{fig:sf_theory_var}
\end{figure}

\subsection{Free-tropospheric temperature change mechanism}
This mechanism is one of the most important in terms of its contribution to the variation of scaling factor across temperature percentiles (Fig. \ref{fig:sf_theory_breakdown}). For all simulations, changes in free-tropospheric temperature lead to suppressed warming of the hottest days (relative to warming of the 50th percentile) because the rate of warming in the free troposphere is generally smaller for hotter days (Fig. \ref{fig:sf_theory_var}a). For the aquaplanet (land) simulations, the suppression of surface warming is greater for the hot (cold) simulations (Fig. \ref{fig:sf_theory_breakdown}b,c). In the free troposphere, the warming between simulations is larger for the aquaplanet compared to land across the entire distribution (Fig. \ref{fig:sf_theory_var}a). This is reflected in a larger absolute scaling factor contribution from this mechanism  for the aquaplanet (deviation of green line from zero in Fig. \ref{fig:sf_theory_breakdown}). However, the difference between aquaplanet and land in terms of surface warming associated with this mechanism is less than that implied by the difference in free-tropospheric warming alone (difference between the dashed and solid lines in Fig. \ref{fig:sf_theory_var}a is $\sim0.5$ whereas the difference in green lines between the top and bottom rows of Fig. \ref{fig:sf_theory_breakdown} is $\sim0.2$). This is partly due to the larger sensitivity of this mechanism for the land simulations (Fig. \ref{fig:sensitivity}a) but primarily due to the lapse rate effect described previously: greater warming in the free troposphere results in a greater reduction in lapse rate (Fig. \ref{fig:lapse_rate}).

\subsection{Hot-get-hotter mechanism}
This mechanism is the weakest in terms of driving differential warming across percentiles (Fig. \ref{fig:sf_theory_breakdown}). For the colder simulations (Fig. \ref{fig:sf_theory_breakdown}a,c), the contribution to the scaling factor is negligible because we average over two regimes: the $\kappa=0.6, 0.8$  simulations are in the ``cold-get-hotter'' regime while the $\kappa=1.0, 1.5$ are in the ``hot-get-hotter'' regime (Fig. \ref{fig:sensitivity}b). For the hotter simulations (Fig. \ref{fig:sf_theory_breakdown}b,d), the contribution is larger for land because the surface temperature anomaly between, for example, hot days and the average day is greater over land (Fig. \ref{fig:sf_theory_var}b). The hot-get-hotter mechanism contributes to the greater overall broadening of the land surface temperature distribution in the hot simulations relative to the cold simulations (Fig. \ref{fig:sf_theory_breakdown}c,d).


\subsection{Relative humidity-change mechanism}
The contribution of the relative humidity-change mechanism is negative (i.e., less than 1; see Fig. \ref{fig:sf_theory_breakdown}) except for the cold land simulations. This is because the relative humidity increases with warming across the surface temperature distribution for the aquaplanet and hot land simulations (Fig. \ref{fig:sf_theory_var}c). The effect of this mechanism on the shape of the temperature distribution is to weakly amplify warming of the hottest days in the aquaplanet and hot land simulations (Fig. \ref{fig:sf_theory_breakdown}a,b,d). For the cold land simulations, relative humidity decreases with warming for the majority of the distribution (Fig. \ref{fig:sf_theory_var}c) resulting in amplified warming of the upper quartiles relative to the lower quartiles (Fig. \ref{fig:sf_theory_breakdown}c).  This mechanism counteracts the contribution from the free-tropospheric temperature change mechanism because changes in relative humidity and free-tropospheric temperature are highly correlated (not shown). The net effect of these two mechanisms on the change of surface temperature distribution is small for the cold simulations (Fig. \ref{fig:sf_theory_breakdown}a,c). For the hot simulations (Fig. \ref{fig:sf_theory_breakdown}b,d), the free-tropospheric temperature (relative humidity) change tends to be dominant for the aquaplanet (land) simulations and contributes to the amplified warming of the coldest (hottest) days. A discussion of the correlation between changes in free-tropospheric temperature and relative humidity is provided in section \ref{sect:wtg}.

\subsection{Drier-get-hotter mechanism}
The drier-get-hotter mechanism is negligible for the aquaplanet simulations (Fig. \ref{fig:sf_theory_breakdown}a,b) due to both small relative humidity anomalies (Fig. \ref{fig:sf_theory_var}d) and a small sensitivity parameter (Fig. \ref{fig:sensitivity}d). In the land simulations (Fig. \ref{fig:sf_theory_breakdown}c,d), hot days being relatively dry (Fig. \ref{fig:sf_theory_var}d) drives amplified warming of those days, all else equal [consistent with \citet{byrne_amplified_2021}]. The contribution is larger for cold simulations (Fig. \ref{fig:sf_theory_breakdown}c), despite smaller relative humidity anomalies (Fig. \ref{fig:sf_theory_var}d), because the sensitivity is larger (Fig. \ref{fig:sensitivity}d).

\subsection{CAPE-change mechanism}
The CAPE-change mechanism provides an overall positive contribution to the scaling factor (Fig. \ref{fig:sf_theory_breakdown}) due to the general increase in CAPE with warming across the temperature distribution (Fig. \ref{fig:sf_theory_var}e, except for some of the hot aquaplanet simulations). This result is consistent with previous work using cloud-system resolving and fully-coupled global models to highlight the role of entrainment in driving larger CAPE in a warmer climate \citep{singh_influence_2013,singh2017increasing}. However, understanding the processes governing CAPE increases in the idealized simulations considered here is a topic for future research. For the land simulations the change in CAPE is roughly constant across the temperature distribution (Fig. \ref{fig:sf_theory_var}e), resulting in a weak influence on the shape of the temperature distribution (Fig. \ref{fig:sf_theory_breakdown}c,d). For the aquaplanet simulations there is an increase in CAPE for the coldest days which drives amplified warming of those days. Note that a CAPE variable outputted by the model (Fig. \ref{fig:cape}) does not show this increase for cold days, complicating the physical interpretation (Appendix \ref{appendix:cape}).

To summarize the effects of the five physical mechanisms on the scaling factor: The free-tropospheric temperature change and relative humidity-change mechanisms broadly cancel out such that the structure of the scaling factor across temperature percentiles is then set by the CAPE-change  mechanism for the aquaplanet simulations. For the land simulations, a combination of the drier-get-hotter and hot-get-hotter mechanisms control the scaling factor's structure, with the relative humidity-change mechanism influential above approximately the 60th percentile. Next, the three change mechanisms are examined in detail.

\section{Decomposing the `change' mechanisms contributing to differential warming}
\label{sect:wtg}

\begin{figure}
\center
\includegraphics[width=35pc]{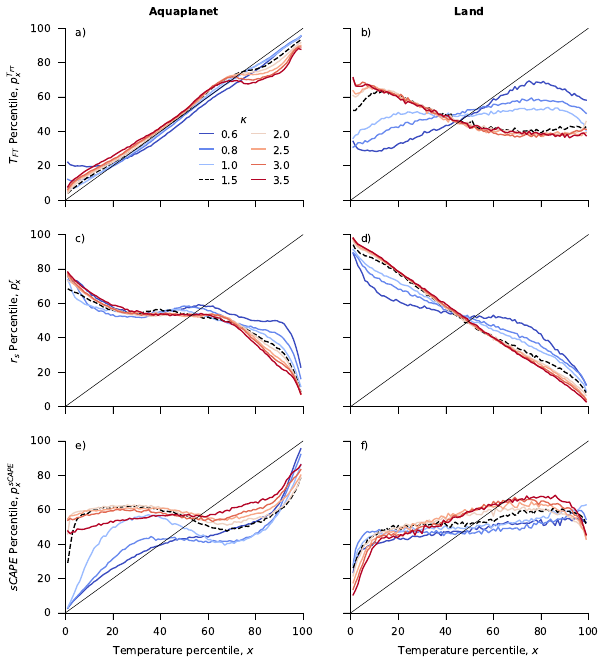}
\caption{Percentile of a given variable $\chi$, denoted $p^{\chi}_x$, corresponding to the average value of $\chi$ conditioned on surface temperature in the percentile range $x-0.5$ to $x+0.5$ versus temperature percentile for the aquaplanet and land simulations. Each row of panels corresponds to a different variable: (a, b) free-tropospheric temperature; (c, d) surface relative humidity; and (e, f) simple CAPE. Solid black lines show the one-to-one relationships.}\label{fig:percentile}
\end{figure}



Three of the mechanisms capable of driving differential warming involve changes in the distributions of physical quantities conditioned on temperature, specifically changes in free-tropospheric temperature, surface relative humidity, and (simple) CAPE [see (\ref{eqn:sf_theory})]. In general, changes in the distribution of a given quantity conditioned on temperature can arise due to: (i) a change in the distribution of that quantity itself; and (ii) a change in the mapping between percentiles of that quantity and temperature percentiles. In this section the effects of each of these contributions to differential warming are decomposed for the three `change' mechanisms. We focus on the free-tropospheric temperature and relative humidity mechanisms, given their importance for differential warming and a prominent correlation between the terms across temperature percentiles (Fig. \ref{fig:sf_theory_breakdown}).

\begin{figure}
\center
\includegraphics[width=35pc]{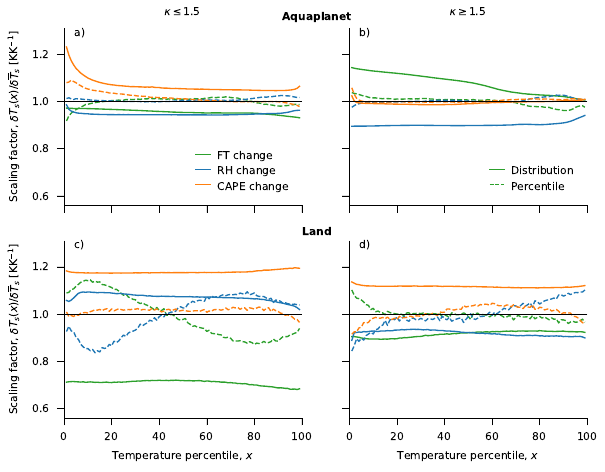}
\caption{Decomposition of the free-tropospheric temperature (green), relative humidity (blue), and CAPE (orange) change contributions to the scaling factor into distribution (solid lines) and percentile components (dashed lines) following (\ref{eqn:p_x_change}). For each variable, the sum of the distribution and percentile components approximately equals the overall scaling factor contribution (see Fig. \ref{fig:sf_theory_breakdown}). The decompositions for the four coldest (a,c) and five warmest (b,d) aquaplanet and land simulations are shown. Note that 1 has been added to all contributions except the free-tropospheric temperature distribution component.}\label{fig:sf_percentile}
\end{figure}


According to the theory (\ref{eqn:sf_theory}), it is not the change in the distribution of a given quantity $\chi$ itself that drives differential warming, but rather the change in that quantity conditioned on the $x$th percentile of surface temperature, i.e. $\delta \chi[x]$. To relate changes in the $\chi$ distribution to changes in $\chi$ conditioned on temperature, we define $p_x^{\chi}$ to be the percentile of $\chi$ corresponding to the average value of $\chi$ conditioned on surface temperature percentiles between $x-0.5$ and $x+0.5$, i.e. $\chi[x]\equiv\chi(p^{\chi}_x)$. Figure  \ref{fig:percentile} shows that, in general, $p^{\chi}_x$ differs substantially from $x$. This implies, for example, that on an aquaplanet hot temperatures in the free troposphere broadly correspond to hot surface temperatures (Fig. \ref{fig:percentile}a), but this is not the case for land temperatures (Fig. \ref{fig:percentile}b). Similarly hot days tend to be relatively dry (i.e., they correspond to a low $p_x^r$) for both aquaplanet and land simulations, and becoming relatively drier as climate warms (Fig. \ref{fig:percentile}c,d).


Using $p^{\chi}_x$, we can approximate changes in the $\chi$ distribution conditioned on surface temperature, $\delta \chi[x]$, as (see Appendix \ref{appendix:percentile} for details):

\begin{equation} \label{eqn:p_x_change}
\delta \chi[x] \approx \delta \chi(p^{\chi}_x) + \left[\chi(p^{\chi}_x+\delta p^{\chi}_x) - \chi(p^{\chi}_x)\right],
\end{equation}
where $\delta \chi(p^{\chi}_x)$ is defined as the \emph{distribution} term, quantifying how the distribution of $\chi$ itself varies with warming. The second term on the RHS of (\ref{eqn:p_x_change}) is defined as the \emph{percentile} term. This term quantifies the effect of the change with warming in the percentile of $\chi$ corresponding to a given percentile of surface temperature, i.e. $\delta p_x^{\chi}$, assuming no change in the distribution of $\chi$ itself. 


Substituting (\ref{eqn:p_x_change}) into (\ref{eqn:sf_theory}) for $\chi \in \{T_{FT}, r_s, sCAPE\}$ results in an alternative version of the scaling factor theory:

\begin{equation} \label{eqn:sf_theory_p_x}
\begin{split}
\frac{\delta T_s(x)}{\delta\overline{T}_s} &\approx
\gamma_{\delta T_{FT}} \left[\frac{\delta T_{FT}(p_x^{T_{FT}})}{\delta \overline{T}_s}
+ \frac{T_{FT}(p_x^{T_{FT}} + \delta p_x^{T_{FT}}) - T_{FT}(p_x^{T_{FT}})}{\delta \overline{T}_s} \right] 
+ \gamma_{\Delta T_s}\frac{\Delta T_s(x)}{\overline{T}_s} \\
&- \gamma_{\delta r} \frac{\overline{T}_s}{\overline{r}_s} 
\left[\frac{\delta r_s(p_x^r)}{\delta \overline{T}_s}
+ \frac{r(p_x^r + \delta p_x^r) - r(p_x^r)}{\delta \overline{T}_s}\right]
- \gamma_{\Delta r} \frac{\Delta r_s[x]}{\overline{r}_s} \\
&+ \frac{\gamma_{\delta T_{FT}}}{R^{\dagger}}\left[\frac{\delta sCAPE(p_x^{sCAPE})}{\delta \overline{T}_s} + \frac{sCAPE(p_x^{sCAPE} + \delta p_x^{sCAPE}) - sCAPE(p_x^{sCAPE})}{\delta \overline{T}_s} \right].
\end{split}
\end{equation}
This version of the theory performs very similarly to (\ref{eqn:sf_theory}) (cf. dashed and dotted lines in Fig. \ref{fig:sf_theory}). The decomposition of each of the `change' mechanisms into distribution and percentile components is shown in Figure \ref{fig:sf_percentile}. 


For the land simulations, the distribution components of the free-tropospheric temperature and CAPE-change mechanisms show little variation across temperature percentile (see solid green and orange lines in Fig. \ref{fig:sf_percentile}c,d). The variations across $x$ shown by these mechanisms in Figure \ref{fig:sf_theory_breakdown}c,d are therefore primarily due to the percentile components. This implies that changes with warming in the shapes of the free-tropospheric temperature and CAPE distributions do not strongly influence the shape of the temperature distribution. But the correspondence between percentiles of surface temperature and percentiles of free-tropospheric temperature and CAPE does change, influencing the response of the surface temperature distribution to warming (Fig. \ref{fig:sf_percentile}c,d).




For the aquaplanet simulations, particularly in hot climates, a narrowing of the free-tropospheric temperature distribution contributes to a narrowing of the surface temperature distribution with warming (Fig. \ref{fig:sf_percentile}b). A narrowing of the free-tropospheric temperature distribution is consistent with \citet{quan_weakening_2025}, who showed weaker free-tropospheric temperature gradients in warmer climates. The CAPE-change scaling factor contribution shows little variation across temperature percentiles except for the coldest days, which are not convectively coupled across all simulations, complicating the interpretation of simple CAPE (see Appendix \ref{appendix:cape} for a discussion). The percentile relative humidity-change component has a smaller magnitude than for the land simulations, despite the percentile changes being of a similar magnitude (Fig. \ref{fig:percentile}c,d). This is because the relative humidity distribution is much broader over land (e.g., for the $\kappa=1.0$ land simulation, the interquartile range of $r_s$ is 16.8$\%$ for land compared to 3.3$\%$ for the aquaplanet), so a given change in percentile results in a greater absolute change in relative humidity. The CAPE distribution is similarly much broader over land (interquartile ranges of 298J kg$^{-1}$ and 53J kg$^{-1}$ for land and aquaplanet, respectively) while the free-tropospheric temperature distribution is pretty similar for both (interquartile ranges of 3.9K and 3.6K for land and aquaplanet, respectively). 




The free-tropospheric temperature and relative humidity-change mechanisms show a strong anti-correlation in their percentile components across $x$, particularly for land (Fig. \ref{fig:sf_percentile}c,d).
To gain intuition into why this correlation exists for the land simulations, consider a control scenario with a moist adiabatic temperature profile (i.e., no CAPE) corresponding to a given surface temperature and relative humidity. As climate warms, hot land days become relatively drier ($\delta p_x^r < 0$; see Fig. \ref{fig:percentile}d). As a result, the temperature profile adjusts to a new moist adiabat with a higher LCL. For fixed surface temperature this surface drying implies a cooling in the free troposphere (see dashed lines in Fig. \ref{fig:schematic}d). This thought experiment suggests a qualitative physical explanation for why a correlation exists between relative surface drying ($\delta p_x^r <0$) and relative cooling of the free troposphere ($\delta p_x^{T_{FT}} <0$) over land, which translates into an anti-correlation in the effects of these mechanisms on the scaling factor (Fig. \ref{fig:sf_percentile}c,d). Here we have taken a ``bottom up'' perspective to interpret how changes in surface relative humidity affect free-tropospheric temperature. But an equally valid approach would be to consider how a relative cooling in the free troposphere affects surface humidity; understanding the direction of causality in this argument is an open question for future work.

\section{Summary and discussion}
\label{sect:summary}



Motivated by recent work demonstrating a strong influence of convection on surface temperatures in a changing climate, here we have developed a detailed theoretical framework for understanding how different physical mechanisms influence the shape of the tropical temperature distribution. Four of these mechanisms capable of changing the shape of the temperature distribution arise from changes in moist adiabatic lapse rates with warming, which are rooted in well-established thermodynamics. First, warming in the free troposphere is reflected at the surface but with reduced magnitude (free-tropospheric temperature change mechanism). Second, a reduction in relative humidity with warming leads to amplified surface warming compared to a case with constant relative humidity (relative humidity-change mechanism). Third, warming is amplified for days that are climatologically hot (hot-get-hotter mechanism) or dry (drier-get-hotter mechanism). An additional mechanism is introduced to account for scenarios where the temperature profile deviates from a moist adiabat (CAPE-change mechanism). 


Applied to the summertime tropics in idealized simulations over a wide range of climates, this framework captures the amplified warming of hot days over land and the suppressed warming of hot days over ocean. The main driver of this contrasting behavior over land versus ocean is that hot days are relatively dry over land compared to the average day (drier-get-hotter mechanism), consistent with results from fully-coupled models \citep{byrne_amplified_2021}. The simulations show little change in the shapes of the free-tropospheric temperature or CAPE distributions with warming. However, the relationship between percentiles of free-tropospheric temperature and surface temperature does shift with warming, with hot days at the surface become relatively cooler in the free troposphere. In isolation, this relative cooling of the free troposphere would be expected to suppress surface warming following the free-tropospheric temperature change mechanism. But the effect of this mechanism is largely canceled by hot days becoming relatively drier; whether this cancellation is a feature of more comprehensive simulations of climate change is a topic for future research. The greater the cancellation between these mechanisms, the more dominant the climatological mechanisms (hot-get-hotter and drier-get-hotter) are in determining changes in the surface temperature distribution, with potential implications for constraining future projections.

The theory developed here is broadly applicable where convection is active, so a natural next step would be to apply it to mid-latitude land regions in summertime. But free-tropospheric temperature gradients in mid-latitudes are not constrained to be weak, so the convection-based theory would likely need to be coupled to the large-scale circulation in some way. Further investigation of the role of CAPE in shaping surface temperatures would also be an interesting avenue for future study, perhaps by combining the framework developed here with theories for thermal stratification \citep[e.g.,][]{singh_influence_2013}. Finally, using the theory to understand the processes driving spatial variability in the response of tropical extreme temperatures to a warming climate \citep[e.g.,][]{huntingford_acceleration_2024} would be another interesting application of this framework.  



\clearpage
\acknowledgments The authors thank Talia Tamarin-Brodsky and Funing Li for helpful discussions. JAMD is supported by a St Leonard's College Doctoral Scholarship awarded by the University of St Andrews and MPB is supported by the UKRI Frontier Research Guarantee scheme (grant number EP/Y027868/1).

%
%
\datastatement 
Code is available on Zenodo: \url{https://zenodo.org/records/15363076}.


\begin{appendices}
\counterwithin{figure}{section}
\counterwithin{equation}{section}

\section{Simple estimate of CAPE}
\label{appendix:cape}
The convective available potential energy \citep[CAPE; e.g. see equation 18.60 in][]{vallis_atmospheric_2017} is found by integrating the temperature deviation between the moist adiabat in the SCE limit, $T_{SCE}$, and the environmental profile, $T$, from the level of free convection (LFC) up to the level of neutral buoyancy (LNB):

\begin{equation} \label{eqn:cape_vallis}
CAPE = -R\int_{p_{LFC}}^{p_{LNB}}(T_{SCE}-T)d\ln p.
\end{equation}
To get from (\ref{eqn:cape_vallis}) to equation (\ref{eqn:cape}) in the main text, we remove the temperature deviation from the integral and approximate that on average it is equal to half its value at the chosen free-tropospheric level, i.e. $\frac{1}{2}(T_{FT,SCE} - T_{FT})$. We also change the integration limits to be between the surface and free-tropospheric level and obtain (\ref{eqn:cape}), our simple CAPE-like metric: $sCAPE = R^{\dagger}(T_{FT,SCE} - T_{FT})$.


\begin{figure}
\center
\includegraphics[width=35pc]{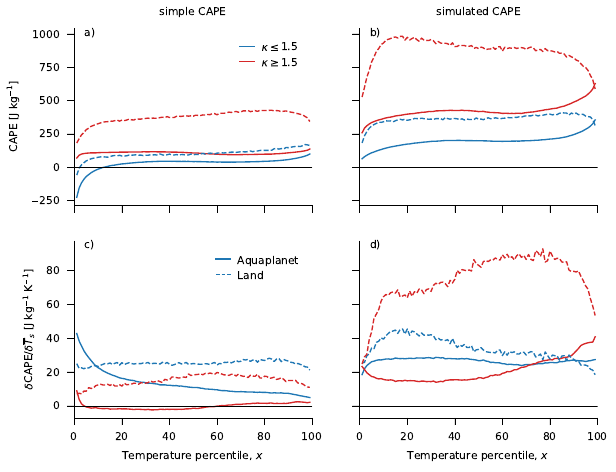}
\caption{Comparison of a simple CAPE-like quantity [$sCAPE$; see (\ref{eqn:cape})] with simulated CAPE. Panels (a) and (b) show $sCAPE$ and simulated CAPE for the cold (blue) and warm (red) simulations. 
Panel (c) is the same as Fig. \ref{fig:sf_theory_var}e but with different $y$-axis limits and panel (d) shows the rate of change of model output CAPE as a function of surface temperature percentile, $x$. Data for the aquaplanet (land) simulations are shown by solid (dashed) lines. In each case, regressions are performed to show a single line for the four coldest (blue) and five hottest (red)
simulations: $\overline{T}_s\chi[x]$ against $\overline{T}_s$ for (a) and (b); whereas $\chi[x]$ against $\overline{T}_s$ for (c) and (d), for $\chi \in \{sCAPE, CAPE\}$.}\label{fig:cape}
\end{figure}

A comparison of panels (a) and (b) in Figure \ref{fig:cape} shows that that our simple CAPE-like quantity, $sCAPE$, has a similar variation across the surface temperature distribution to simulated CAPE: both show a slight increase with temperature for the aquaplanet simulations but are roughly temperature independent for land. Both metrics both show larger CAPE over land and in the hotter simulations.

There are more substantial difference in how the CAPE metrics respond to warming (Fig. \ref{fig:cape}c,d). Our simple CAPE-like quantity increases at a faster rate for the colder simulations (Fig. \ref{fig:cape}c) but for the land simulations, the simulated CAPE increase is greater for the hotter simulations (Fig. \ref{fig:cape}d). This is likely because in the hotter simulations, the level to which convection reaches rises more with warming. The cold aquaplanet simulations also show the biggest increase in simple CAPE for the coldest days, which is not found for simulated CAPE. This may be related to the negative values in Figure \ref{fig:cape}a for the coldest days of the $\kappa\leq 1.5$ simulations. This indicates that $T_{FT}>T_{FT,SCE}$ and thus the level of neutral buoyancy is closer to the surface than our chosen free tropospheric level, i.e. convection does not reach the level in the atmosphere where simple CAPE is calculated. So the large increase is likely reflecting the fact that with warming, convection reaches higher in the atmosphere (or a transition towards a convection dominated vertical coupling regime). Both of these discrepancies are a consequence of estimating simple CAPE using a single level in the atmosphere.


\section{Conversion between percentiles}
\label{appendix:percentile}
The theory for the scaling factor theory (\ref{eqn:sf_theory}) is formulated in terms of variables $\chi \in \{T_{FT}, r_s, CAPE\}$ and their changes with warming, $\delta \chi$, conditioned on the $x^{th}$ percentile of surface temperature: $\chi[x]$ and $\delta \chi[x]$. To relate these changes to the actual distributions, $\chi(x)$, following \citet{byrne_amplified_2021} a quantity $p_x^{\chi}$ is introduced and defined to be the percentile of $\chi$ corresponding to the average value of $\chi$ conditioned on surface temperature percentile $x$, i.e. $\chi[x] \equiv \chi(p_x^{\chi})$. The relationship between $\delta \chi[x]$ and $\delta \chi(p_x^{\chi})$ is derived below for the change from a cold to a hot climate:

\begin{eqnarray}
    \delta \chi[x] &=& \chi^{hot}[x] - \chi^{cold}[x] \label{eqn:percentile_deriv1} \\
    &=& \chi^{hot}(p^{\chi,hot}_x) - \chi^{cold}(p^{\chi,cold}_x) \label{eqn:percentile_deriv2} \\
    &=& \chi^{hot}(p^{\chi}_x + \delta p^{\chi}_x) - \chi^{cold}(p^{\chi}_x) \label{eqn:percentile_deriv3} \\ 
    &=& \left[\chi^{hot}(p^{\chi}_x) - \chi^{cold}(p^{\chi}_x)\right] + \left[\chi^{hot}(p^{\chi}_x + \delta p^{\chi}_x) - \chi^{hot}(p^{\chi}_x)\right] \label{eqn:percentile_deriv4} \\
    &=& \delta \chi(p_x^{\chi}) + \left[\chi(p^{\chi}_x + \delta p^{\chi}_x) - \chi(p^{\chi}_x)\right] + \left[\delta \chi(p^{\chi}_x + \delta p^{\chi}_x) - \delta \chi(p^{\chi}_x)\right],
    \label{eqn:percentile_deriv5}
\end{eqnarray}
where we set $p^{\chi,cold}_x = p^{\chi}_x$ going from (\ref{eqn:percentile_deriv2}) to (\ref{eqn:percentile_deriv3}) and set $\chi^{cold} = \chi$ going from (\ref{eqn:percentile_deriv4}) to (\ref{eqn:percentile_deriv5}). 
To get from (\ref{eqn:percentile_deriv5}) to (\ref{eqn:p_x_change}), the small non-linear contributions are neglected [i.e., the $\delta \chi(p^{\chi}_x + \delta p^{\chi}_x) - \delta \chi(p^{\chi}_x)$ term].

\section{Theory derivation}
\label{appendix:derivation}

\subsection{Moist static energy}

\noindent To derive a theory for the sensitivity of the $x$th percentile of (near-)surface temperature, we start by relating the surface MSE, $h_s$, to the free-tropospheric MSE at saturation, $h_{FT}^*$. In general these quantities will differ by an amount $\epsilon$, which is zero by definition in the SCE limit:

\begin{equation} \label{eqn:deriv_epsilon}
h_s = h_{FT}^* + \epsilon.
\end{equation}

\subsection{Relating $z_{FT}$ to $T_{FT}$}

Equation (\ref{eqn:deriv_epsilon}) involves two related free-tropospheric quantities, $T_{FT}$ and $z_{FT}$. When considering changes with warming, we would like to combine the free-tropospheric changes in temperature and geopotential height into a single term. Following \citet{zhang_upper_2023} we use hydrostatic balance and the ideal gas law to write:

\begin{equation} \label{eqn:deriv_z1}
d\ln p = -\frac{g}{RT(p)}dz
\end{equation}
and define a bulk lapse rate, $\Gamma$, between the surface and pressure level $p$ in the free troposphere such that

\begin{equation} \label{eqn:deriv_z2}
z(p) - z_s = \frac{T_s - T(p)}{\Gamma}.
\end{equation}
Combining (\ref{eqn:deriv_z1}) and (\ref{eqn:deriv_z2}), we obtain:

\begin{equation} \label{eqn:deriv_z3}
d\ln p = \frac{g}{R\Gamma}d\ln T.
\end{equation}
Integrating between the surface and free troposphere results in:

\begin{equation} \label{eqn:deriv_z4}
\ln\left(\frac{p_s}{p_{FT}}\right) = \frac{g}{R\Gamma}\ln\left(\frac{T_s}{T_{FT}}\right).
\end{equation}
We can rearrange (\ref{eqn:deriv_z4}) to write an expression for bulk lapse rate:

\begin{equation} \label{eqn:deriv_z_lapse}
\frac{1}{\Gamma} = \frac{2R^{\dagger}}{g\ln\left(\frac{T_s}{T_{FT}}\right)},
\end{equation}
where we have introduced for convenience a modified gas constant $R^{\dagger} \equiv R\ln \left(\frac{p_s}{p_{FT}}\right)/2$. We can re-write the denominator in the form $\ln(1+y)$, with $y = \frac{T_s-T_{FT}}{T_{FT}}$. Using the first two terms in the Taylor expansion $\frac{1}{\ln(1+y)} \approx \frac{1}{y} + \frac{1}{2}$, we approximate (\ref{eqn:deriv_z_lapse}) as:

\begin{equation} \label{eqn:deriv_z_lapse_taylor}
\frac{1}{\Gamma} \approx \frac{2R^{\dagger}}{g}\left(\frac{T_{FT}}{T_s-T_{FT}} + \frac{1}{2}\right).
\end{equation}
We substitute (\ref{eqn:deriv_z_lapse_taylor}) into (\ref{eqn:deriv_z2}) at $p=p_{FT}$ to obtain our final relation between $z_{FT}$ and $T_{FT}$:

\begin{equation} \label{eqn:deriv_z_final}
z_{FT} - z_s = \frac{R^{\dagger}}{g}(T_s + T_{FT}) + A_z,
\end{equation}
where $A_z$ is a term representing the error in this relation. We will use similar terms (denoted with the symbol $A$) throughout the derivation to quantify the importance of each approximation in the final theoretical scaling factor. The inclusion of these terms makes all the equations in the following derivation exact. Apart from the two $A_{\beta}$ terms which are dimensionless, all error terms have the same units as MSE (J kg$^{-1}$).

\subsection{Modified MSE}

\noindent Using (\ref{eqn:deriv_z_final}), we obtain an equation for the saturation free-tropospheric MSE that no longer depends explicitly on $z_{FT}$:

\begin{equation} \label{eqn:deriv_mse_ft_no_z}
h_{FT}^* = (c_p+R^{\dagger})T_{FT} + L_v q^*_{FT} + gz_s + R^{\dagger}T_s + A_z.
\end{equation}
This motivates the introduction of a modified MSE, $h^{\dagger}$, independent of surface variables:

\begin{equation} \label{eqn:deriv_mse_mod_ft}
h^{\dagger} = h^*_{FT} - R^{\dagger}T_s - gz_s = \left(c_p + R^{\dagger}\right) T_{FT} + L_v q^*_{FT} + A_z.
\end{equation}
Combining (\ref{eqn:deriv_epsilon}) and (\ref{eqn:deriv_mse_mod_ft}) gives a second expression for $h^{\dagger}$ in terms of surface variables and $\epsilon$:

\begin{equation} \label{eqn:deriv_mse_mod_s}
h^{\dagger} = \left(c_p - R^{\dagger}\right)T_s + L_v q_s - \epsilon.
\end{equation}

\subsection{Change in the modified MSE anomaly with warming, $\delta \Delta h^{\dagger}$ - Free Troposphere}
The goal here is to derive an equation for $\delta T_s(x)$ for a given change in mean surface temperature, $\delta \overline{T_s}$, where $x$ is surface temperature percentile. Given the two expressions for $h^{\dagger}$, (\ref{eqn:deriv_mse_mod_ft}) and (\ref{eqn:deriv_mse_mod_s}), this can be achieved by equating two different expressions for $\delta \left(h^{\dagger}[x]-\overline{h}^{\dagger}\right)$. In this section, we obtain an expression from free-tropospheric temperature (\ref{eqn:deriv_mse_mod_ft}) while in the next section we obtain an expression from surface temperature and humidity (\ref{eqn:deriv_mse_mod_s}).
%

Performing a Taylor expansion of $h^{\dagger}[x]$ about a reference values of $T_{FT}$, we find:

\begin{eqnarray}
h^{\dagger}[x] &=& \left(c_p+R^{\dagger}\right) \tilde{T}_{FT} + L_v \tilde{q}^*_{FT} + \left(c_p + R^{\dagger} + L_v \tilde{\alpha}_{FT} \tilde{q}^*_{FT}\right) \Delta T_{FT}[x] \nonumber \\
&+& A_{FT\Delta}[x] + A_z[x]
\label{eqn:deriv_mse_taylor}\\
\Delta h^{\dagger}[x] &=& \tilde{\beta}_{FT1} \Delta T_{FT}[x] + A_{FT\Delta}[x] + \Delta A_z[x],
\label{eqn:deriv_mse_anom}
\end{eqnarray}
where here and throughout the derivation, $\Delta \chi = \chi[x] - \tilde{\chi}$ for a given variable $\chi$ and $\tilde{\chi}$ indicates a reference value of $\chi$ we are free to choose [we will make specific choices for reference values at end of the derivation in order to get to derive (\ref{eqn:sf_theory})]. The parameter $\beta_{FT1} = \frac{dh^{\dagger}}{dT_{FT}} = c_p + R^{\dagger} + L_v \alpha_{FT} q^*_{FT}$ depends on the Clausius-Clapeyron parameter $\alpha$ in the free troposphere, defined as the fractional change of saturation specific humidity following a 1 K temperature change \citep{held2006robust}, i.e. $\alpha \equiv (dq^*/dT) / q^*$.

In (\ref{eqn:deriv_mse_taylor}) we introduced the \emph{approximate} term $A_{FT\Delta}=\sum_{n=2}^{\infty}\frac{1}{n!}\widetilde{\frac{\partial^n h^{\dagger}}{\partial T_{FT}^n}}(\Delta T_{FT})^n$, grouping together higher order terms in the Taylor expansion. We introduce a further dimensionless \emph{approximate} term, $\tilde{A}_{FT\beta}$, in order to quantify the change in $\tilde{\beta}_{FT1}$ with warming between simulations (indicated by $\delta$):

\begin{equation} \label{eqn:deriv_beta_ft_change}
\delta \tilde{\beta}_{FT1} = \tilde{\beta}_{FT2}(1 + \tilde{A}_{FT\beta})\frac{\delta \tilde{T}_{FT}}{\tilde{T}_{FT}},
\end{equation}
where $\beta_{FT2}\frac{\delta T_{FT}}{T_{FT}} = \frac{d^2h^{\dagger}}{dT_{FT}^2} \delta T_{FT} = \frac{d\beta_{FT1}}{dT_{FT}} \delta T_{FT} = L_v \alpha_{FT} q^*_{FT}(\alpha_{FT} T_{FT} - 2)\frac{\delta T_{FT}}{T_{FT}}$ is the linear Taylor expansion term. Using (\ref{eqn:deriv_beta_ft_change}), we obtain an expression for the change in modified MSE anomaly:

\begin{eqnarray}
\delta \Delta h^{\dagger}[x] &=& \left(\tilde{\beta}_{FT1} + \delta \tilde{\beta}_{FT1}\right)\delta \Delta T_{FT}[x] + \delta \tilde{\beta}_{FT1}\Delta T_{FT}[x] + \delta A_{FT\Delta}[x] + \delta \Delta A_z[x]\label{eqn:deriv_mse_mod_ft_anom_change0}\\
&=& \left(\tilde{\beta}_{FT1} + \delta \tilde{\beta}_{FT1}\right)\delta \Delta T_{FT}[x] + \tilde{\beta}_{FT2}(1 + \tilde{A}_{FT\beta}) 
\left(\Delta T_{FT}[x] + \frac{\Delta \epsilon[x]}{\tilde{\beta}_{FT1}}\right) \frac{\delta \tilde{T}_{FT}}{\tilde{T}_{FT}}- \nonumber \\ 
&\quad& \frac{\delta \tilde{\beta}_{FT1}}{\tilde{\beta}_{FT1}}\Delta \epsilon[x] + \delta A_{FT\Delta}[x] + \delta \Delta A_z[x],
\label{eqn:deriv_mse_mod_ft_anom_change1}
\end{eqnarray}
where in (\ref{eqn:deriv_mse_mod_ft_anom_change1}) we have separated out the $\epsilon$ terms for ease of introducing the sCAPE variable later.

\subsubsection*{Replacing $\Delta T_{FT}$}

To isolate physical mechanisms associated with climatological surface temperature and relative humidity anomalies, we aim to write $\Delta T_{FT} = f(\Delta T_s, \Delta r_s, \Delta \epsilon)$. As a first step, we invert (\ref{eqn:deriv_mse_anom}) to give $\tilde{\beta}_{FT1} \Delta T_{FT}[x] = \Delta h^{\dagger}[x] - A_{FT\Delta}[x] - \Delta A_z[x]$. Performing a Taylor expansion of (\ref{eqn:deriv_mse_mod_s}) about the mean, we obtain:

\begin{eqnarray} \label{eqn:deriv_mse_mod_s_anom_taylor0}
\Delta h^{\dagger}[x] &=& \tilde{\beta}_{s1}\left(1 + \tilde{\mu}\frac{\Delta r_s[x]}{\tilde{r}_s}\right)\Delta T_s[x] + L_v \tilde{q}_s\frac{\Delta r_s[x]}{\tilde{r}_s} - \Delta \epsilon[x] + A_{s\Delta}[x] \\
&=& \Delta h^{\dagger}_0[x] - \Delta \epsilon[x] + A_{s\Delta}[x],\label{eqn:deriv_mse_mod_s_anom_taylor}
\end{eqnarray}
where in an analogous way to $\beta_{FT1}$ we have introduced $\beta_{s1} = \frac{\partial h^{\dagger}}{\partial T_s} = c_p - R^{\dagger} + L_v \alpha_sq_s$ to quantify the sensitivity of modified MSE to changes in surface temperature. We also introduce the parameter $\mu = 1 - \frac{c_p - R^{\dagger}}{c_p - R^{\dagger} + L_v \alpha_s q_s} = \frac{L_v \alpha_s q_s}{\beta_{s1}}$ to quantify the importance of the non-linear term such that $\frac{\partial^2 h^{\dagger}}{\partial T_s \partial r_s} = \frac{\beta_{s1}\mu}{r_s}$. The modified MSE anomaly due to linear temperature and relative humidity contributions only is contained within $\Delta h^{\dagger}_0$. The \emph{approximate} term $A_{s\Delta}$ is analogous to $A_{FT\Delta}$ in that it quantifies the higher order terms in the Taylor expansion:

\begin{equation} \label{eqn:deriv_approx_s_anom}
A_{s\Delta}[x] = \left(1 + \frac{\Delta r_s[x]}{\tilde{r}_s}\right)\sum_{n=2}^{\infty}\frac{1}{n!}\widetilde{\frac{\partial^n h^{\dagger}}{\partial T_{s}^n}}(\Delta T_{s})^n
= \left(1 + \frac{\Delta r_s[x]}{\tilde{r}_s}\right)A_{s\Delta T}[x].
\end{equation}

Combining (\ref{eqn:deriv_mse_anom}), (\ref{eqn:deriv_mse_mod_ft_anom_change1}) and (\ref{eqn:deriv_mse_mod_s_anom_taylor}), we obtain:

\begin{align} \label{eqn:deriv_mse_mod_ft_anom_change2}
\begin{split}
\delta \Delta h^{\dagger}[x] &= \frac{\tilde{\beta}_{FT2}}{\tilde{\beta}_{FT1}}(1 + \tilde{A}_{FT\beta}) 
\left(\Delta h^{\dagger}_0[x] + A_{s\Delta}[x] - A_{FT\Delta}[x] - \Delta A_z[x]\right) \frac{\delta \tilde{T}_{FT}}{\tilde{T}_{FT}} \\ 
&+ (\tilde{\beta}_{FT1} + \delta \tilde{\beta}_{FT1})\delta \Delta T_{FT}[x]- \frac{\delta \tilde{\beta}_{FT1}}{\tilde{\beta}_{FT1}}\Delta \epsilon[x] + \delta A_{FT\Delta}[x] + \delta \Delta A_z[x].
\end{split}
\end{align}

\subsubsection*{Replacing $\delta \tilde{T}_{FT}$}
We would like to relate $\delta \tilde{T}_{FT}$ to the change in surface variables with warming, because in the final scaling factor we want to quantify the average difference between simulations using surface rather than free-tropospheric temperature. Taking changes with warming of (\ref{eqn:deriv_mse_mod_ft}) for the reference day, we obtain:

\begin{equation} \label{eqn:deriv_mse_mean_change_ft}
\delta \tilde{h}^{\dagger} = \tilde{\beta}_{FT1} \delta \tilde{T}_{FT} + \tilde{A}_{FT\delta} + \delta \tilde{A}_z,
\end{equation}
where $\tilde{A}_{FT\delta} = \sum_{n=2}^{\infty}\frac{1}{n!}\widetilde{\frac{\partial^n h^{\dagger}}{\partial T_{FT}^n}}(\delta T_{FT})^n$ is the \emph{approximate} term containing the non-linear Taylor expansion contributions.

Similarly taking changes with warming of (\ref{eqn:deriv_mse_mod_s}) for the reference day, we obtain:

\begin{align} \label{eqn:deriv_mse_mean_change_s}
\begin{split}
\delta \tilde{h}^{\dagger} &= \tilde{\beta}_{s1}\left(1 + \tilde{\mu}\frac{\delta \tilde{r}_s}{\tilde{r}_s}\right)\delta \tilde{T}_s
+ L_v \tilde{q}_s\frac{\delta \tilde{r}_s}{\tilde{r}_s} - \delta \tilde{\epsilon} + \tilde{A}_{s\delta}\\
&= \delta \tilde{h}_0^{\dagger} - \delta \tilde{\epsilon} + \tilde{A}_{s\delta},
\end{split}
\end{align}
where $\delta \tilde{h}_0$ is the linear Taylor expansion change from relative humidity and temperature changes only, similar to (\ref{eqn:deriv_approx_s_anom}), $\tilde{A}_{s\delta} = \left(1 + \frac{\delta r_s}{\tilde{r}_s}\right)\sum_{n=2}^{\infty}\frac{1}{n!}\widetilde{\frac{\partial^n h^{\dagger}}{\partial T_{s}^n}}(\delta \tilde{T}_{s})^n$. 

Combining (\ref{eqn:deriv_mse_mean_change_ft}) and (\ref{eqn:deriv_mse_mean_change_s}), we get an equation for $\delta \tilde{T}_{FT}$ in terms of surface quantities:

\begin{equation} \label{eqn:deriv_temp_ft_mean_change}
\tilde{\beta}_{FT1} \delta \tilde{T}_{FT} = \delta \tilde{h}_0^{\dagger} - \delta \tilde{\epsilon} + \tilde{A}_{s\delta} - \tilde{A}_{FT\delta} - \delta \tilde{A}_z.
\end{equation}
Substituting (\ref{eqn:deriv_temp_ft_mean_change}) into (\ref{eqn:deriv_mse_mod_ft_anom_change2}), we arrive at our final equation for the free-tropospheric $\delta \Delta h^{\dagger}[x]$:

\begin{align} \label{eqn:deriv_mse_mod_ft_anom_change_final0}
\begin{split}
\delta \Delta h^{\dagger}[x] &= \frac{\tilde{\beta}_{FT2}}{\tilde{\beta}_{FT1}}(1 + \tilde{A}_{FT\beta}) 
\left(\Delta h^{\dagger}_0[x] + A_{s\Delta}[x] - A_{FT\Delta}[x] - \Delta A_z[x]\right) \frac{\delta \tilde{h}^{\dagger}_0-\delta \tilde{\epsilon}+\tilde{A}_{s\delta} - \tilde{A}_{FT\delta} - \delta \tilde{A}_z}{\tilde{\beta}_{FT1}\tilde{T}_{FT}} \\ 
&+ \tilde{\beta}_{FT1}\delta T_{FT}[x] - \delta \tilde{h}_0^{\dagger}  
+ \delta \tilde{\beta}_{FT1}\delta \Delta T_{FT}[x] + \delta \tilde{\epsilon}
- \frac{\delta \tilde{\beta}_{FT1}}{\tilde{\beta}_{FT1}}\Delta \epsilon[x] \\
&- \tilde{A}_{s\delta} + \tilde{A}_{FT\delta} + \delta \tilde{A}_z + \delta A_{FT\Delta}[x] + \delta \Delta A_z[x]
\end{split}
\end{align}
To simplify this expression, we gather approximations and small non-linear contributions together based on what is responsible for the error:

\begin{align} \label{eqn:deriv_mse_mod_ft_anom_change_final}
\begin{split}
\delta \Delta h^{\dagger}[x] &= \frac{\tilde{\beta}_{FT2}}{\tilde{\beta}_{FT1}} 
\left(\tilde{\beta}_{s1}\Delta T_s[x]
+ L_v \tilde{q}_s\frac{\Delta r_s[x]}{\tilde{r}_s}\right) \frac{\delta \tilde{h}^{\dagger}_0-\delta \tilde{\epsilon}}{\tilde{\beta}_{FT1}\tilde{T}_{FT}} + \\ 
&\quad \tilde{\beta}_{FT1}\delta T_{FT}[x] - \delta \tilde{h}_0^{\dagger}  
 + \delta \tilde{\epsilon}
- \frac{\delta \tilde{\beta}_{FT1}}{\tilde{\beta}_{FT1}}\Delta \epsilon[x] + \\
&\quad A_{\delta \Delta T_{FT}}[x] + \delta \Delta A_{z}[x] + A_{\Delta 1}[x] + A_{\Delta T\Delta r1}[x] + \tilde{A}_{\delta} + A_{NL}[x],
\end{split}
\end{align}
where:

\begin{eqnarray}
A_{\delta \Delta T_{FT}} &=& \delta A_{FT\Delta}[x] + \delta \tilde{\beta}_{FT1}\delta \Delta T_{FT}[x] \\
A_{\Delta 1}[x] &=& \frac{\tilde{\beta}_{FT2}}{\tilde{\beta}_{FT1}} 
\frac{\delta \tilde{h}^{\dagger}_0 - \delta \tilde{\epsilon}}{\tilde{\beta}_{FT1}\tilde{T}_{FT}}
(A_{s\Delta}[x] - A_{FT\Delta}[x] - \Delta A_{z}[x]) +\nonumber \\
&\quad&  \frac{\tilde{\beta}_{FT2}}{\tilde{\beta}_{FT1}}
\frac{\Delta h^{\dagger}_0[x] - \Delta \epsilon[x]}{\tilde{\beta}_{FT1}\tilde{T}_{FT}}
\left(\left(\delta \tilde{h}^{\dagger}_0 - \delta \tilde{\epsilon}\right) \tilde{A}_{FT\beta} 
+ \left(1 + \tilde{A}_{FT\beta}\right)
\left(\tilde{A}_{s\delta} - \tilde{A}_{FT\delta} - \delta \tilde{A}_{z}\right)
\right) \\
A_{\Delta T\Delta r1}[x] &=& \frac{\tilde{\beta}_{FT2}}{\tilde{\beta}_{FT1}} \frac{\delta \tilde{h}^{\dagger}_0 - \delta \tilde{\epsilon}}{\tilde{\beta}_{FT1}\tilde{T}_{FT}}
\tilde{\mu} \tilde{\beta}_{s1}\frac{\Delta r_s[x]}{\tilde{r}_s}\Delta T_s[x] \\
\tilde{A}_{\delta} &=& -\left(\tilde{A}_{s\delta} - \tilde{A}_{FT\delta} - \delta \tilde{A}_{z}\right) \\
A_{NL}[x] &=& \frac{\tilde{\beta}_{FT2}}{\tilde{\beta}_{FT1}} \frac{1}{\tilde{\beta}_{FT1}\tilde{T}_{FT}} 
\left(\left(\delta \tilde{h}^{\dagger}_0 - \delta \tilde{\epsilon}\right) \tilde{A}_{FT\beta}
+ \left(1 + \tilde{A}_{FT\beta}\right)\left(\tilde{A}_{s\delta} - \tilde{A}_{FT\delta} - \delta \tilde{A}_{z}\right)\right) \\
&\quad& (A_{s\Delta}[x] - A_{FT\Delta}[x] - \Delta A_z[x]).
\end{eqnarray}

\subsection{Change in modified MSE anomaly with warming, $\delta \Delta h^{\dagger}$ - Surface}

Taking the change of (\ref{eqn:deriv_mse_mod_s_anom_taylor0}) with warming and using $\delta (\tilde{\mu}\tilde{\beta}_{s1}) = \delta \tilde{\beta}_{s1}$, we obtain:

\begin{align} \label{eqn:deriv_mse_mod_anom_surf_change0}
\begin{split}
\delta \Delta h^{\dagger}[x] &= \delta \tilde{\beta}_{s1}\left[1 + \frac{\Delta r_s[x]}{\tilde{r}_s} 
+ \delta \left(\frac{r_s[x]}{\tilde{r}_s}\right)\right]\left(\Delta T_s[x] + \delta \Delta T_s[x]\right) + \\
&\quad \tilde{\beta}_{s1}\left[\tilde{\mu}\delta \left(\frac{r_s[x]}{\tilde{r}_s}\right)\left(\Delta T_s[x] + \delta \Delta T_s[x]\right) + \left(1 + \tilde{\mu}\frac{\Delta r_s[x]}{\tilde{r}_s}\right)\delta \Delta T_s[x]\right] + \\
&\quad \delta (L_v \tilde{q}_s)\left[\delta \left(\frac{r_s[x]}{\tilde{r}_s}\right) + \frac{\Delta r_s[x]}{\tilde{r}_s}\right] + L_v \tilde{q}_s \delta \left(\frac{r_s[x]}{\tilde{r}_s}\right) - \delta \Delta \epsilon[x] + \\
&\quad \left(1 + \frac{\Delta r_s[x]}{\tilde{r}_s}\right)\delta A_{s\Delta T}[x] +
 \delta \left(\frac{r_s[x]}{\tilde{r}_s}\right) (A_{s\Delta T}[x] + \delta A_{s\Delta T}[x]),
\end{split}
\end{align}
where in the final line, we used (\ref{eqn:deriv_approx_s_anom}) to replace $\delta A_{s\Delta}[x]$ with $A_{s\Delta T}[x]$ terms. If we again gather the non-linear and \emph{approximate} terms, we can write this as:

\begin{align} \label{eqn:deriv_mse_mod_anom_surf_change1}
\begin{split}
\delta \Delta h^{\dagger}[x] &= \tilde{\beta}_{s1} \delta \Delta T_s[x] + L_v\tilde{q}_s \delta \left(\frac{r_s[x]}{\tilde{r}_s}\right)
+ \Delta T_s[x] \delta \tilde{\beta}_{s1} + \frac{\Delta r_s[x]}{\tilde{r}_s} \delta(L_v \tilde{q}_s)
- \delta \Delta \epsilon[x] +  \\
&\quad A_{\delta \Delta T_s}[x] + A_{\delta r1}[x] + A_{\delta \Delta T_s\delta r}[x] + A_{\Delta T\Delta r2},
\end{split}
\end{align}
where:

\begin{eqnarray}
A_{\delta \Delta T_s}[x] &=& \left(1 + \frac{\Delta r_s[x]}{\tilde{r}_s}\right)\delta A_{s\Delta T}[x] + \left[\tilde{\mu}\tilde{\beta}_{s1} \frac{\Delta r_s[x]}{\tilde{r}_s}
+ \left(1 + \frac{\Delta r_s[x]}{\tilde{r}_s}\right) \delta \tilde{\beta}_{s1}\right]\delta \Delta T_s[x] \\
A_{\delta r1}[x] &=& \left(A_{s\Delta T}[x] + \delta(L_v \tilde{q}_s) + (\tilde{\mu}\tilde{\beta}_{s1} + \delta \tilde{\beta}_{s1})\Delta T_s[x]\right)
\delta \left(\frac{r_s[x]}{\tilde{r}_s}\right)\\
A_{\delta \Delta T_s\delta r}[x] &=& \delta A_{s\Delta T}[x] \delta \left(\frac{r_s[x]}{\tilde{r}_s}\right)
+ (\tilde{\mu}\tilde{\beta}_{s1} + \delta \tilde{\beta}_{s1})\delta \left(\frac{r_s[x]}{\tilde{r}_s}\right)\delta \Delta T_s[x] \\
A_{\Delta T\Delta r2}[x] &=& \delta \tilde{\beta}_{s1}\frac{\Delta r_s[x]}{\tilde{r}_s} \Delta T_s[x].
\end{eqnarray}

To replace $\delta \tilde{\beta}_{s1}$, we use an analogous equation to (\ref{eqn:deriv_beta_ft_change}) but account for the relative humidity dependence of $\tilde{\beta}_{s1}$:

\begin{equation} \label{eqn:deriv_beta_s_change}
\delta \tilde{\beta}_{s1} = \tilde{\beta}_{s2}(1 + \tilde{A}_{s\beta}) \left(1 + \frac{\delta \tilde{r}_s}{\tilde{r}_s}\right)
\frac{\delta \tilde{T}_s}{\tilde{T}_s} + \tilde{\mu}\tilde{\beta}_{s1} \frac{\delta \tilde{r}_s}{\tilde{r}_s},
\end{equation}
where $\beta_{s2} = T_s \frac{\partial^2h^{\dagger}}{\partial T_s^2} =  T_s\frac{\partial \beta_{s1}}{\partial T_s} = L_v \alpha_s q_s(\alpha_s T_s - 2)$ and $\tilde{A}_{s\beta}$ is the dimensionless \emph{approximate} term. We would also like to replace $\delta \left(\frac{r_s[x]}{\tilde{r}_s}\right)$ with $\delta r_s[x]$, which we can do with the following:

\begin{equation} \label{eqn:deriv_r_frac_change}
L_v\tilde{q}_s \delta \left(\frac{r_s[x]}{\tilde{r}_s}\right) = L_v\tilde{q}_s \left[\frac{\delta r_s[x]}{\tilde{r}_s}
- \left(1+\frac{\Delta r_s[x]}{\tilde{r}_s}\right)\frac{\delta \tilde{r}_s}{\tilde{r}_s}\right] + A_{\delta r2}[x],
\end{equation}
where we have introduced the \emph{approximate} term $A_{\delta r2}$ to store the higher order terms in this expansion (note that $A_{\delta r2}=0$ if $\delta \tilde{r}_s=0$). Lastly, to replace $\delta (L_v\tilde{q}_s)$, we combine (\ref{eqn:deriv_mse_mod_s}) with (\ref{eqn:deriv_mse_mean_change_s}):

\begin{eqnarray} \label{eqn:deriv_sphum_change}
\delta (L_v \tilde{q}_s) &=& \delta \left(\tilde{h}^{\dagger} + \tilde{\epsilon} - (c_p-R^{\dagger})\tilde{T}_s\right) \nonumber \\
&=& \tilde{\mu}\tilde{\beta}_{s1}\left(1 + \frac{\delta \tilde{r}_s}{\tilde{r}_s} \right)\delta \tilde{T}_s
+ L_v\tilde{q}_s \frac{\delta \tilde{r}_s}{\tilde{r}_s} + \tilde{A}_{s\delta},
\end{eqnarray}
where we used $\tilde{\mu}\tilde{\beta}_{s1} = L_v\tilde{\alpha}_s \tilde{q}_s = \tilde{\beta}_{s1} - (c_p-R^{\dagger})\tilde{T}_s$. 

Combining (\ref{eqn:deriv_mse_mod_anom_surf_change1}), (\ref{eqn:deriv_beta_s_change}),  (\ref{eqn:deriv_r_frac_change}) and (\ref{eqn:deriv_sphum_change}) we obtain our final expression for the surface derived $\delta \Delta h^{\dagger}$:

\begin{align} \label{eqn:deriv_mse_mod_anom_surf_change_final}
\begin{split}
\delta \Delta h^{\dagger}[x] &= L_v\tilde{q}_s\frac{\delta \Delta r_s[x]}{\tilde{r}_s} + \tilde{\beta}_{s1} \delta \Delta T_s[x]
 - \delta \Delta \epsilon[x] + \\
&\quad \Delta T_s[x]
\left(\tilde{\beta}_{s2} \left(1 + \frac{\delta \tilde{r}_s}{\tilde{r}_s}\right)
\frac{\delta \tilde{T}_s}{\tilde{T}_s} + \tilde{\mu}\tilde{\beta}_{s1} \frac{\delta \tilde{r}_s}{\tilde{r}_s}\right) + \\
&\quad \frac{\Delta r_s[x]}{\tilde{r}_s}
\tilde{\mu}\tilde{\beta}_{s1}\left(1 + \frac{\delta \tilde{r}_s}{\tilde{r}_s} \right)\delta \tilde{T}_s + \\
&\quad A_{\delta \Delta T_s}[x] + A_{\delta r}[x] + A_{\delta \Delta T_s\delta r}[x] + A_{\Delta T\Delta r2}[x] + A_{\Delta 2}[x],
\end{split}
\end{align}
where $A_{\delta r}[x] = A_{\delta r1}[x] + A_{\delta r2}[x]$ and we have introduced another \emph{approximate} term, $A_{\Delta 2}$, which arises due to the $\tilde{A}_{s\beta}$ and $\tilde{A}_{s\delta}$ introduced from (\ref{eqn:deriv_beta_s_change}) and (\ref{eqn:deriv_sphum_change}) respectively:

\begin{equation} \label{eqn:deriv_approx_anom2}
A_{\Delta 2}[x] =  \tilde{A}_{s\beta} \tilde{\beta}_{s2} \left(1 + \frac{\delta \tilde{r}_s}{\tilde{r}_s}\right)
\frac{\delta \tilde{T}_s}{\tilde{T}_s} \Delta T_s[x] + \tilde{A}_{s\delta} \frac{\Delta r_s[x]}{\tilde{r}_s}.
\end{equation}

\subsection{Replacing $\epsilon$ with sCAPE}
To get an estimate for the scaling factor, we equate (\ref{eqn:deriv_mse_mod_ft_anom_change_final}) with (\ref{eqn:deriv_mse_mod_anom_surf_change_final}). On doing this, we end up with a contribution involving $\epsilon$ terms: $\delta \Delta \epsilon[x] + \delta \tilde{\epsilon} - \frac{\delta \tilde{\beta}_{FT1}}{\tilde{\beta}_{FT1}}\Delta \epsilon[x]$. We would like to express this in terms of simple CAPE, defined through (\ref{eqn:cape}), instead of $\epsilon$. 

To do this, we first define $T_{FT,SCE}$ to be $T_{FT}(T_s, r_s, \epsilon=0, A_z)$ i.e. the solution of $h^{\dagger}_{SCE}=(c_p-R^{\dagger})T_s + L_v q_s = (c_p + R^{\dagger})T_{FT,SCE} + L_vq^*(T_{FT,SCE}) + A_z$. 

\subsubsection*{Conditioned on $x$}
\noindent Analogous to (\ref{eqn:deriv_mse_anom}), we have:

\begin{equation} \label{eqn:deriv_approx_sce_x}
h^{\dagger}_{SCE}[x] - \tilde{h}^{\dagger}
= \tilde{\beta}_{FT1}(T_{FT,SCE}[x]-\tilde{T}_{FT}) + A_{FT\Delta,SCE}[x] + \Delta A_z[x],
\end{equation}
where $A_{FT\Delta,SCE}[x]=\sum_{n=2}^{\infty}\frac{1}{n!}\widetilde{\frac{\partial^n h^{\dagger}}{\partial T_{FT}^n}}(T_{FT,SCE}[x]-\tilde{T}_{FT})^n$ is an \emph{approximate} term similar to $A_{FT\Delta}[x]$. Then using (\ref{eqn:deriv_mse_mod_s}) and (\ref{eqn:deriv_mse_anom}) for both the general and SCE cases, we get:

\begin{eqnarray} 
(c_p-R^{\dagger})T_s[x] + L_v q_s[x] - \epsilon[x] - \tilde{h}^{\dagger}
&=& \tilde{\beta}_{FT1}(T_{FT}[x]-\tilde{T}_{FT}) + A_{FT\Delta}[x] + \Delta A_z[x]  \label{eqn:deriv_cape_x1} \\ 
(c_p-R^{\dagger})T_s[x] + L_v q_s[x] - \tilde{h}^{\dagger}&=& \tilde{\beta}_{FT1}(T_{FT,SCE}[x]-\tilde{T}_{FT}) + \nonumber \\ &\quad& 
A_{FT\Delta,SCE}[x] + \Delta A_z[x]. \label{eqn:deriv_cape_x2}
\end{eqnarray}
Subtracting (\ref{eqn:deriv_cape_x1}) from (\ref{eqn:deriv_cape_x2}) and using the definition of simple CAPE (\ref{eqn:cape}), we obtain:

\begin{equation} \label{eqn:deriv_epsilon_cape_x}
\epsilon[x] = \frac{\tilde{\beta}_{FT1}}{R^{\dagger}}sCAPE[x] + A_{FT\Delta,SCE}[x]-A_{FT\Delta}[x].
\end{equation}

\subsubsection*{Reference CAPE}
\noindent Similar to (\ref{eqn:deriv_cape_x2}) but for the reference case, we have:

\begin{equation} \label{eqn:deriv_cape_ref1}
(c_p-R^{\dagger})\tilde{T}_s + L_v \tilde{q}_s - \tilde{h}^{\dagger} = \tilde{\beta}_{FT1}(\tilde{T}_{FT,SCE}-\tilde{T}_{FT}) + \tilde{A}_{FT\Delta,SCE},
\end{equation}
where $\tilde{A}_{FT\Delta,SCE} = \sum_{n=2}^{\infty}\frac{1}{n!}\widetilde{\frac{\partial^n h^{\dagger}}{\partial T_{FT}^n}}(\tilde{T}_{FT,SCE} - \tilde{T}_{FT})^n$. Then, substituting in (\ref{eqn:deriv_mse_mod_s}) we obtain:

\begin{equation} \label{eqn:deriv_epsilon_cape_ref}
\tilde{\epsilon} = \frac{\tilde{\beta}_{FT1}}{R^{\dagger}}\widetilde{sCAPE} + \tilde{A}_{FT\Delta,SCE}.
\end{equation}
The change with warming is given by:

\begin{equation} \label{eqn:deriv_epsilon_cape_ref_change}
\delta \tilde{\epsilon} = \frac{\tilde{\beta}_{FT1}+\delta \tilde{\beta}_{FT1}}{R^{\dagger}}\delta \widetilde{sCAPE} + \frac{\delta \tilde{\beta}_{FT1}}{R^{\dagger}}\widetilde{sCAPE} + \delta \tilde{A}_{FT\Delta,SCE}.
\end{equation}

\subsubsection*{CAPE Anomaly}
\noindent Combining (\ref{eqn:deriv_epsilon_cape_x}) and (\ref{eqn:deriv_epsilon_cape_ref}), we have:

\begin{equation} \label{eqn:deriv_epsilon_cape_anom}
\Delta \epsilon[x] = \frac{\tilde{\beta}_{FT1}}{R^{\dagger}}\Delta sCAPE[x] + A_{FT\Delta,SCE}[x] - \tilde{A}_{FT\Delta,SCE} -A_{FT\Delta}[x]
\end{equation}
and the change with warming is given by:

\begin{align} \label{eqn:deriv_epsilon_cape_anom_change}
\begin{split}
\delta \Delta \epsilon[x] &= \frac{\tilde{\beta}_{FT1}+\delta \tilde{\beta}_{FT1}}{R^{\dagger}}\delta \Delta sCAPE[x] + \frac{\delta \tilde{\beta}_{FT1}}{R^{\dagger}}\Delta sCAPE[x] + \\
&\quad \delta (A_{FT\Delta,SCE}[x] - \tilde{A}_{FT\Delta,SCE} -A_{FT\Delta}[x]).
\end{split}
\end{align}

\subsubsection*{Total CAPE contribution}

\noindent Combining (\ref{eqn:deriv_epsilon_cape_ref_change}), (\ref{eqn:deriv_epsilon_cape_anom}) and (\ref{eqn:deriv_epsilon_cape_anom_change}) gives:

\begin{equation} \label{eqn:deriv_epsilon_cape_sf_cont}
\delta \Delta \epsilon[x] + \delta \tilde{\epsilon} - \frac{\delta \tilde{\beta}_{FT1}}{\tilde{\beta}_{FT1}}\Delta \epsilon[x]
= \frac{\tilde{\beta}_{FT1}}{R^{\dagger}}\delta sCAPE[x] + A_{CAPE}[x],
\end{equation}
where we have combined all approximations as well as the non-linear term into a new \emph{approximate} term:

\begin{align} \label{eqn:deriv_cape_approx}
\begin{split}
A_{CAPE}[x] &= \frac{\delta \tilde{\beta}_{FT1}}{R^{\dagger}}\left(\widetilde{sCAPE} + \delta sCAPE[x]\right)
+ \delta (A_{FT\Delta,SCE}[x]-A_{FT\Delta}[x]) -  \\
&\quad \frac{\delta \tilde{\beta}_{FT1}}{\tilde{\beta}_{FT1}}\left(A_{FT\Delta,SCE}[x] - \tilde{A}_{FT\Delta,SCE} -A_{FT\Delta}[x] \right).
\end{split}
\end{align}

\subsection{Temperature change equation, $\delta T_s(x)$}
On equating (\ref{eqn:deriv_mse_mod_ft_anom_change_final}) with (\ref{eqn:deriv_mse_mod_anom_surf_change_final}), and using (\ref{eqn:deriv_epsilon_cape_sf_cont}), we arrive at an equation for $\delta T_s(x)$:

\begin{align} \label{eqn:deriv_temp_surf_change}
\begin{split}
\tilde{\beta}_{s1} \delta T_s[x] &= \tilde{\beta}_{s1}\delta \tilde{T}_s - \delta \tilde{h}_0^{\dagger}
+ \tilde{\beta}_{FT1}\delta T_{FT}[x] -L_v\tilde{q}_s\frac{\delta \Delta r_s[x]}{\tilde{r}_s}
+ \frac{\tilde{\beta}_{FT1}}{R^{\dagger}}\delta sCAPE[x] +  \\
&\quad \Delta T_s \left[\frac{\tilde{\beta}_{FT2}}{\tilde{\beta}_{FT1}} \frac{\tilde{\beta}_{s1}}{\tilde{\beta}_{FT1}} \frac{\delta \tilde{h}_0^{\dagger}-\delta \tilde{\epsilon}}{\tilde{T}_{FT}}
- \tilde{\beta}_{s2} \left(1 + \frac{\delta \tilde{r}_s}{\tilde{r}_s}\right)
\frac{\delta \tilde{T}_s}{\tilde{T}_s} - \tilde{\mu}\tilde{\beta}_{s1} \frac{\delta \tilde{r}_s}{\tilde{r}_s}\right] - \\
&\quad \frac{\Delta r_s[x]}{\tilde{r}_s}\left[\tilde{\mu}\tilde{\beta}_{s1}\left(1 + \frac{\delta \tilde{r}_s}{\tilde{r}_s} \right)\delta \tilde{T}_s - \frac{\tilde{\beta}_{FT2}}{\tilde{\beta}_{FT1}} \frac{\delta \tilde{h}_0^{\dagger}-\delta \tilde{\epsilon}}{\tilde{\beta}_{FT1}\tilde{T}_{FT}}L_v \tilde{q}_s\right] + \\
&\quad A_{\delta \Delta T_{FT}}[x] + \delta \Delta A_z[x] + A_{\Delta}[x] + A_{\Delta T\Delta r}[x] + \tilde{A}_{\delta} + A_{NL}[x] +\\
&\quad A_{CAPE}[x] - A_{\delta \Delta T_s}[x] - A_{\delta r}[x] - A_{\delta \Delta T_s \delta r}[x],
\end{split}
\end{align}
where $A_{\Delta}[x] = A_{\Delta 1}[x] - A_{\Delta 2}[x]$ and $A_{\Delta T \Delta r}[x] = A_{\Delta T \Delta r 1}[x] - A_{\Delta T \Delta r 2}[x]$.

\subsection{Choice of reference variables}
The equation (\ref{eqn:deriv_temp_surf_change}) is clearly more complex than (\ref{eqn:sf_theory}). This is because it is more general; choices for the reference variables have not yet been made. The choices we make to obtain (\ref{eqn:sf_theory}) are:

\begin{tabular}{ll}  
    $\bullet$ $\tilde{T}_s = \overline{T}_s$ & $\bullet$ $\delta \tilde{T}_s = \delta \overline{T}_s$ \\
    $\bullet$ $\tilde{r}_s = \overline{r}_s$ & $\bullet$ $\delta \tilde{r}_s = 0$ \\
    $\bullet$ $\tilde{\epsilon} = 0$ & $\bullet$ $\delta \tilde{\epsilon} = 0$ \\
    $\bullet$ $\tilde{A}_z = 0$ & $\bullet$ $\delta \tilde{A}_z = 0$.
\end{tabular}

From these four variables, $\tilde{T}_{FT}$ can be computed by equating (\ref{eqn:deriv_mse_mod_ft}) and (\ref{eqn:deriv_mse_mod_s}). Given our choices, $\tilde{T}_{FT}$ (and also $\tilde{\beta}_{FT1}$ and $\tilde{\beta}_{FT2}$) is uniquely determined by mean surface temperature and relative humidity.

\subsection{Sensitivity $\gamma$ parameters}

Substituting in the reference variables into (\ref{eqn:deriv_temp_surf_change}) and neglecting the \emph{approximate} terms results in (\ref{eqn:sf_theory}) after dividing by $\tilde{\beta}_{s1}\delta \tilde{T}_s$ [note that with the reference choices, $\delta \tilde{h}_0^{\dagger} = \tilde{\beta}_{s1}\delta \tilde{T}_s$ using (\ref{eqn:deriv_mse_mean_change_s})]. The dimensionless sensitivity parameters, depending only on $\overline{T}_s$ and $\overline{r}_s$, are then given by:

\begin{eqnarray}
\gamma_{\delta T_{FT}} &=& \frac{\tilde{\beta}_{FT1}}{\tilde{\beta}_{s1}} \label{eqn:gamma_ft} \\
\gamma_{\delta r} &=& \frac{L_v\tilde{q}_{s}}{\tilde{\beta}_{s1}\tilde{T}_s} \label{eqn:gamma_r_change} \\
\gamma_{\Delta T_s} &=&  \frac{\tilde{\beta}_{FT2}}{\tilde{\beta}_{FT1}}\frac{\tilde{\beta}_{s1}}{\tilde{\beta}_{FT1}}\frac{\tilde{T}_{s}}{\tilde{T}_{FT}} - \frac{\tilde{\beta}_{s2}}{\tilde{\beta}_{s1}} \label{eqn:gamma_temp}  \\
\gamma_{\Delta r} &=&  \tilde{\mu} - \frac{\tilde{\beta}_{FT2}}{\tilde{\beta}_{FT1}}\frac{L_v\tilde{q}_s}{\tilde{\beta}_{FT1}\tilde{T}_{FT}}.\label{eqn:gamma_r_anom} 
\end{eqnarray}

\subsection{Approximate terms}
\begin{figure}
\center
\includegraphics[width=35pc]{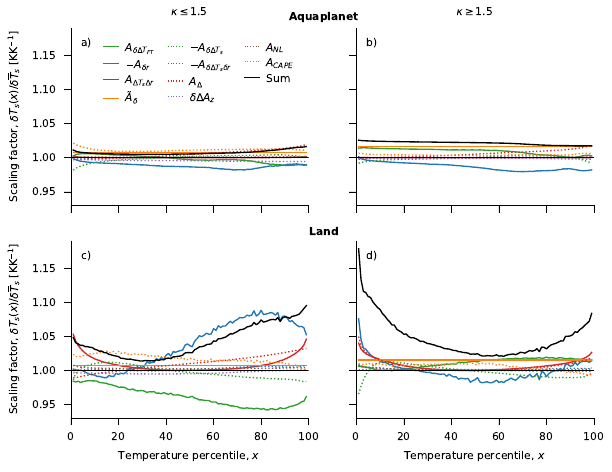}
\caption{Approximations neglected in deriving the theoretical scaling factor (\ref{eqn:sf_theory}) for the aquaplanet (a, b) and land simulations (c, d). The sum of all approximation terms (black line), along with the five contributions shown in Fig. \ref{fig:sf_theory_breakdown}, equal the simulated scaling factors (solid lines in Fig. \ref{eqn:sf_theory}) exactly. Solid lines are used to emphasize the most substantial approximations. Each term has had $1$ added to it. Note the different $y$-axis limits compared to Fig. \ref{fig:sf_theory_breakdown}.}\label{fig:sf_approx}
\end{figure}

The \emph{approximate} terms for the idealized simulations are shown in Figure \ref{fig:sf_approx}. The terms are larger for the land simulations and the most substantial are $A_{\delta \Delta T_{FT}}[x]$ (green solid line), $A_{\delta r}[x]$ (blue solid line), $A_{\Delta T_{s}\Delta r_{s}}[x]$ (red solid line) and $\tilde{A}_{\delta}$ (orange solid line). The inclusion of these terms to the theoretical scaling factor would better capture the behavior at low and high percentiles (Fig. \ref{fig:sf_theory}). So to improve on the theoretical scaling factor for this case, one may want to consider deriving simplified expressions for these four terms.

We have grouped the \emph{approximate} terms together with the intention of highlighting reasons whereby the framework of (\ref{eqn:sf_theory}) may not be appropriate to a particular situation. For example, if the climatological anomaly terms are large ($A_{\Delta}$ and $A_{\Delta T_s \Delta r}$) it may be that the season being considered is too large (i.e. too many months), or that the choices of reference values are not appropriate. The other terms all arise due to the change between simulations; if those are large, it may be more appropriate to consider simulations that are more similar (e.g., smaller $\delta \overline{T}_s$).

\subsection{Note on usage}
Equation (\ref{eqn:deriv_temp_surf_change}) is exact under any scenario, but each term other than those involving $\epsilon$ or CAPE quantifies how the moist adiabatic temperature profile changes with warming. So in non-convective scenarios, it is not appropriate because the profile is far from a moist adiabat. For such a scenario, an alternative starting equation to represent vertical coupling would be more appropriate.

\end{appendices}
\clearpage
\bibliographystyle{ametsocV6}
\bibliography{references}

\begin{thebibliography}{62}
\providecommand{\natexlab}[1]{#1}
\providecommand{\url}[1]{\texttt{#1}}
\renewcommand{\UrlFont}{\rmfamily}
\providecommand{\urlprefix}{URL }
\expandafter\ifx\csname urlstyle\endcsname\relax
  \providecommand{\doi}[1]{https://doi.org/\discretionary{}{}{}#1}\else
  \providecommand{\doi}{https://doi.org/\discretionary{}{}{}\begingroup \urlstyle{rm}\Url}\fi
\providecommand{\eprint}[2][]{\url{#2}}

\bibitem[{Berg et~al.(2016)}]{berg_landatmosphere_2016}
Berg, A., and Coauthors, 2016: Land–atmosphere feedbacks amplify aridity increase over land under global warming. \textit{Nature Climate Change}, \textbf{6~(9)}, 869--874, \doi{10.1038/nclimate3029}.

\bibitem[{Boos(2015)}]{boos2015review}
Boos, W.~R., 2015: A review of recent progress on tibet’s role in the south asian monsoon. \textit{CLIVAR Exch}, \textbf{19~(66)}, 23--27.

\bibitem[{Byrne(2021)}]{byrne_amplified_2021}
Byrne, M.~P., 2021: Amplified warming of extreme temperatures over tropical land. \textit{Nature Geoscience}, \textbf{14~(11)}, 837--841, \doi{10.1038/s41561-021-00828-8}.

\bibitem[{Byrne and O'Gorman(2013)Byrne, and O'Gorman}]{byrne_ogorman_2013b}
Byrne, M.~P., and P.~A. O'Gorman, 2013: Link between land-ocean warming contrast and surface relative humidities in simulations with coupled climate models. \textit{Geophysical Research Letters}, \textbf{40~(19)}, 5223--5227.

\bibitem[{Byrne and O’Gorman(2013)Byrne, and O’Gorman}]{byrne_landocean_2013}
Byrne, M.~P., and P.~A. O’Gorman, 2013: Land–{Ocean} {Warming} {Contrast} over a {Wide} {Range} of {Climates}: {Convective} {Quasi}-{Equilibrium} {Theory} and {Idealized} {Simulations}. \textit{Journal of Climate}, \textbf{26~(12)}, 4000--4016, \doi{10.1175/JCLI-D-12-00262.1}.

\bibitem[{Byrne and O’Gorman(2016)Byrne, and O’Gorman}]{byrne_understanding_2016}
Byrne, M.~P., and P.~A. O’Gorman, 2016: Understanding {Decreases} in {Land} {Relative} {Humidity} with {Global} {Warming}: {Conceptual} {Model} and {GCM} {Simulations}. \textit{Journal of Climate}, \textbf{29~(24)}, 9045--9061, \doi{10.1175/JCLI-D-16-0351.1}.

\bibitem[{Byrne and O’Gorman(2018)Byrne, and O’Gorman}]{byrne_trends_2018}
Byrne, M.~P., and P.~A. O’Gorman, 2018: Trends in continental temperature and humidity directly linked to ocean warming. \textit{Proceedings of the National Academy of Sciences}, \textbf{115~(19)}, 4863--4868, \doi{10.1073/pnas.1722312115}.

\bibitem[{Byrne et~al.(2024)}]{byrne_et_al_2024}
Byrne, M.~P., and Coauthors, 2024: Theory and the future of land-climate science. \textit{Nature Geoscience}, \textbf{17~(11)}, 1079--1086.

\bibitem[{Duan et~al.(2024)Duan, Ahmed,, and Neelin}]{duan2024moist}
Duan, S.~Q., F.~Ahmed, and J.~D. Neelin, 2024: Moist heatwaves intensified by entrainment of dry air that limits deep convection. \textit{Nature Geoscience}, \textbf{17}, 837–--844.

\bibitem[{Emanuel(2007)}]{emanuel2007quasi}
Emanuel, K., 2007: Quasi-equilibrium dynamics of the tropical atmosphere. \textit{The Global Circulation of the Atmosphere}, 186--218.

\bibitem[{Emanuel et~al.(1994)Emanuel, David~Neelin,, and Bretherton}]{emanuel_large-scale_1994}
Emanuel, K.~A., J.~David~Neelin, and C.~S. Bretherton, 1994: On large-scale circulations in convecting atmospheres. \textit{Quarterly Journal of the Royal Meteorological Society}, \textbf{120~(519)}, 1111--1143, \doi{10.1002/qj.49712051902}.

\bibitem[{Fischer and Schär(2009)Fischer, and Schär}]{fischer_future_2009}
Fischer, E.~M., and C.~Schär, 2009: Future changes in daily summer temperature variability: driving processes and role for temperature extremes. \textit{Climate Dynamics}, \textbf{33~(7)}, 917--935, \doi{10.1007/s00382-008-0473-8}.

\bibitem[{Fischer et~al.(2007)Fischer, Seneviratne, Vidale, L{\"u}thi,, and Sch{\"a}r}]{fischer2007soil}
Fischer, E.~M., S.~I. Seneviratne, P.~L. Vidale, D.~L{\"u}thi, and C.~Sch{\"a}r, 2007: Soil moisture--atmosphere interactions during the 2003 european summer heat wave. \textit{Journal of Climate}, \textbf{20~(20)}, 5081--5099.

\bibitem[{Flannigan et~al.(2013)Flannigan, Cantin, de~Groot, Wotton, Newbery,, and Gowman}]{flannigan_global_2013}
Flannigan, M., A.~S. Cantin, W.~J. de~Groot, M.~Wotton, A.~Newbery, and L.~M. Gowman, 2013: Global wildland fire season severity in the 21st century. \textit{Forest Ecology and Management}, \textbf{294}, 54--61, \doi{10.1016/j.foreco.2012.10.022}.

\bibitem[{Frierson(2007)}]{frierson_dynamics_2007}
Frierson, D. M.~W., 2007: The {Dynamics} of {Idealized} {Convection} {Schemes} and {Their} {Effect} on the {Zonally} {Averaged} {Tropical} {Circulation}. \textit{Journal of the Atmospheric Sciences}, \textbf{64~(6)}, 1959--1976, \doi{10.1175/JAS3935.1}.

\bibitem[{Frierson et~al.(2006)Frierson, Held,, and Zurita-Gotor}]{frierson_gray-radiation_2006}
Frierson, D. M.~W., I.~M. Held, and P.~Zurita-Gotor, 2006: A {Gray}-{Radiation} {Aquaplanet} {Moist} {GCM}. {Part} {I}: {Static} {Stability} and {Eddy} {Scale}. \textit{Journal of the Atmospheric Sciences}, \textbf{63~(10)}, 2548--2566, \doi{10.1175/JAS3753.1}.

\bibitem[{Frölicher et~al.(2018)Frölicher, Fischer,, and Gruber}]{frolicher_marine_2018}
Frölicher, T.~L., E.~M. Fischer, and N.~Gruber, 2018: Marine heatwaves under global warming. \textit{Nature}, \textbf{560~(7718)}, 360--364, \doi{10.1038/s41586-018-0383-9}.

\bibitem[{Gasparrini and Armstrong(2011)Gasparrini, and Armstrong}]{gasparrini_impact_2011}
Gasparrini, A., and B.~Armstrong, 2011: The {Impact} of {Heat} {Waves} on {Mortality}. \textit{Epidemiology}, \textbf{22~(1)}, 68, \doi{10.1097/EDE.0b013e3181fdcd99}.

\bibitem[{Held and Soden(2006)Held, and Soden}]{held2006robust}
Held, I.~M., and B.~J. Soden, 2006: Robust responses of the hydrological cycle to global warming. \textit{Journal of climate}, \textbf{19~(21)}, 5686--5699.

\bibitem[{Holmes et~al.(2016)Holmes, Woollings, Hawkins,, and de~Vries}]{holmes_robust_2016}
Holmes, C.~R., T.~Woollings, E.~Hawkins, and H.~de~Vries, 2016: Robust {Future} {Changes} in {Temperature} {Variability} under {Greenhouse} {Gas} {Forcing} and the {Relationship} with {Thermal} {Advection}. \textit{Journal of Climate}, \textbf{29~(6)}, 2221--2236, \doi{10.1175/JCLI-D-14-00735.1}.

\bibitem[{Holton(2004)}]{holton_introduction_2004}
Holton, J.~R., 2004: \textit{An introduction to dynamic meteorology}. Fourth edition ed., International geophysics series; {International} geophysics series; {International} geophysics series; {International} geophysics series; v. 88, Elsevier Academic Press, Burlington, MA, section: xii, 535 pages : illustrations, maps ; 24 cm.

\bibitem[{Hughes et~al.(2017)}]{hughes_global_2017}
Hughes, T.~P., and Coauthors, 2017: Global warming and recurrent mass bleaching of corals. \textit{Nature}, \textbf{543~(7645)}, 373--377, \doi{10.1038/nature21707}.

\bibitem[{Huntingford et~al.(2024)Huntingford, Cox, Ritchie, Clarke, Parry,, and Williamson}]{huntingford_acceleration_2024}
Huntingford, C., P.~M. Cox, P.~D.~L. Ritchie, J.~J. Clarke, I.~M. Parry, and M.~S. Williamson, 2024: Acceleration of daily land temperature extremes and correlations with surface energy fluxes. \textit{npj Climate and Atmospheric Science}, \textbf{7~(1)}, 1--10, \doi{10.1038/s41612-024-00626-0}.

\bibitem[{Korty and Schneider(2007)Korty, and Schneider}]{korty_climatology_2007}
Korty, R.~L., and T.~Schneider, 2007: A {Climatology} of the {Tropospheric} {Thermal} {Stratification} {Using} {Saturation} {Potential} {Vorticity}. \textit{Journal of Climate}, \textbf{20~(24)}, 5977--5991, \doi{10.1175/2007JCLI1788.1}.

\bibitem[{Levine and Schneider(2011)Levine, and Schneider}]{levine_response_2011}
Levine, X.~J., and T.~Schneider, 2011: Response of the {Hadley} {Circulation} to {Climate} {Change} in an {Aquaplanet} {GCM} {Coupled} to a {Simple} {Representation} of {Ocean} {Heat} {Transport}. \textit{Journal of the Atmospheric Sciences}, \textbf{68~(4)}, 769--783, \doi{10.1175/2010JAS3553.1}.

\bibitem[{Li and Tamarin-Brodsky(2025)Li, and Tamarin-Brodsky}]{li2025atmospheric}
Li, F., and T.~Tamarin-Brodsky, 2025: Atmospheric stability sets maximum moist heat and convection in the midlatitudes. \textit{arXiv preprint arXiv:2501.05351}.

\bibitem[{Linz et~al.(2020)Linz, Chen, Zhang,, and Zhang}]{linz_framework_2020}
Linz, M., G.~Chen, B.~Zhang, and P.~Zhang, 2020: A {Framework} for {Understanding} {How} {Dynamics} {Shape} {Temperature} {Distributions}. \textit{Geophysical Research Letters}, \textbf{47~(4)}, e2019GL085\,684, \doi{10.1029/2019GL085684}.

\bibitem[{Manabe(1969)}]{manabe1969climate}
Manabe, S., 1969: Climate and the ocean circulation: I. the atmospheric circulation and the hydrology of the earth's surface. \textit{Monthly weather review}, \textbf{97~(11)}, 739--774.

\bibitem[{McKinnon et~al.(2016)McKinnon, Rhines, Tingley,, and Huybers}]{mckinnon_changing_2016}
McKinnon, K.~A., A.~Rhines, M.~P. Tingley, and P.~Huybers, 2016: The changing shape of {Northern} {Hemisphere} summer temperature distributions. \textit{Journal of Geophysical Research: Atmospheres}, \textbf{121~(15)}, 8849--8868, \doi{10.1002/2016JD025292}.

\bibitem[{McKinnon et~al.(2024)McKinnon, Simpson,, and Williams}]{mckinnon_et_al_2024}
McKinnon, K.~A., I.~R. Simpson, and A.~P. Williams, 2024: The pace of change of summertime temperature extremes. \textit{Proceedings of the National Academy of Sciences}, \textbf{121~(42)}, e2406143\,121.

\bibitem[{Miralles et~al.(2014)Miralles, Teuling, van Heerwaarden,, and Vilà-Guerau~de Arellano}]{miralles_mega-heatwave_2014}
Miralles, D.~G., A.~J. Teuling, C.~C. van Heerwaarden, and J.~Vilà-Guerau~de Arellano, 2014: Mega-heatwave temperatures due to combined soil desiccation and atmospheric heat accumulation. \textit{Nature Geoscience}, \textbf{7~(5)}, 345--349, \doi{10.1038/ngeo2141}.

\bibitem[{Nie et~al.(2010)Nie, Boos,, and Kuang}]{nie2010observational}
Nie, J., W.~R. Boos, and Z.~Kuang, 2010: Observational evaluation of a convective quasi-equilibrium view of monsoons. \textit{Journal of Climate}, \textbf{23~(16)}, 4416--4428.

\bibitem[{Oliver et~al.(2018)}]{oliver_longer_2018}
Oliver, E. C.~J., and Coauthors, 2018: Longer and more frequent marine heatwaves over the past century. \textit{Nature Communications}, \textbf{9~(1)}, 1324, \doi{10.1038/s41467-018-03732-9}.

\bibitem[{O’Gorman and Schneider(2008)O’Gorman, and Schneider}]{ogorman_hydrological_2008}
O’Gorman, P.~A., and T.~Schneider, 2008: The {Hydrological} {Cycle} over a {Wide} {Range} of {Climates} {Simulated} with an {Idealized} {GCM}. \textit{Journal of Climate}, \textbf{21~(15)}, 3815 -- 3832, \doi{10.1175/2007JCLI2065.1}.

\bibitem[{O’Gorman and Schneider(2009)O’Gorman, and Schneider}]{ogorman_scaling_2009}
O’Gorman, P.~A., and T.~Schneider, 2009: Scaling of {Precipitation} {Extremes} over a {Wide} {Range} of {Climates} {Simulated} with an {Idealized} {GCM}. \textit{Journal of Climate}, \textbf{22~(21)}, 5676--5685, \doi{10.1175/2009JCLI2701.1}.

\bibitem[{Patterson(2023)}]{patterson_north-west_2023}
Patterson, M., 2023: North-{West} {Europe} {Hottest} {Days} {Are} {Warming} {Twice} as {Fast} as {Mean} {Summer} {Days}. \textit{Geophysical Research Letters}, \textbf{50~(10)}, e2023GL102\,757, \doi{10.1029/2023GL102757}.

\bibitem[{Perkins(2015)}]{perkins_review_2015}
Perkins, S.~E., 2015: A review on the scientific understanding of heatwaves—{Their} measurement, driving mechanisms, and changes at the global scale. \textit{Atmospheric Research}, \textbf{164-165}, 242--267, \doi{10.1016/j.atmosres.2015.05.014}.

\bibitem[{Pietschnig et~al.(2021)Pietschnig, Swann, Lambert,, and Vallis}]{pietschnig_response_2021}
Pietschnig, M., A.~L.~S. Swann, F.~H. Lambert, and G.~K. Vallis, 2021: Response of {Tropical} {Rainfall} to {Reduced} {Evapotranspiration} {Depends} on {Continental} {Extent}. \textit{Journal of Climate}, \textbf{34~(23)}, 9221--9234, \doi{10.1175/JCLI-D-21-0195.1}.

\bibitem[{Pithan and Mauritsen(2014)Pithan, and Mauritsen}]{pithan_arctic_2014}
Pithan, F., and T.~Mauritsen, 2014: Arctic amplification dominated by temperature feedbacks in contemporary climate models. \textit{Nature Geoscience}, \textbf{7~(3)}, 181--184, \doi{10.1038/ngeo2071}.

\bibitem[{Quan et~al.(2025)Quan, Zhang,, and Fueglistaler}]{quan_weakening_2025}
Quan, H., Y.~Zhang, and S.~Fueglistaler, 2025: Weakening of {Tropical} {Free}-{Tropospheric} {Temperature} {Gradients} with {Global} {Warming}. \textit{Journal of the Atmospheric Sciences}, \textbf{82~(1)}, 31--43, \doi{10.1175/JAS-D-24-0140.1}.

\bibitem[{Romps(2017)}]{romps_exact_2017}
Romps, D.~M., 2017: Exact {Expression} for the {Lifting} {Condensation} {Level}. \textit{Journal of the Atmospheric Sciences}, \textbf{74~(12)}, 3891--3900, \doi{10.1175/JAS-D-17-0102.1}.

\bibitem[{R{\"o}thlisberger and Papritz(2023)R{\"o}thlisberger, and Papritz}]{rothlisberger2023quantifying}
R{\"o}thlisberger, M., and L.~Papritz, 2023: Quantifying the physical processes leading to atmospheric hot extremes at a global scale. \textit{Nature Geoscience}, \textbf{16~(3)}, 210--216.

\bibitem[{Röthlisberger and Papritz(2023)Röthlisberger, and Papritz}]{rothlisberger_quantifying_2023}
Röthlisberger, M., and L.~Papritz, 2023: Quantifying the physical processes leading to atmospheric hot extremes at a global scale. \textit{Nature Geoscience}, \textbf{16~(3)}, 210--216, \doi{10.1038/s41561-023-01126-1}.

\bibitem[{Schneider et~al.(2015)Schneider, Bischoff,, and Płotka}]{schneider_physics_2015}
Schneider, T., T.~Bischoff, and H.~Płotka, 2015: Physics of {Changes} in {Synoptic} {Midlatitude} {Temperature} {Variability}. \textit{Journal of Climate}, \textbf{28~(6)}, 2312--2331, \doi{10.1175/JCLI-D-14-00632.1}.

\bibitem[{Screen(2014)}]{screen_arctic_2014}
Screen, J.~A., 2014: Arctic amplification decreases temperature variance in northern mid- to high-latitudes. \textit{Nature Climate Change}, \textbf{4~(7)}, 577--582, \doi{10.1038/nclimate2268}.

\bibitem[{Seneviratne et~al.(2010)Seneviratne, Corti, Davin, Hirschi, Jaeger, Lehner, Orlowsky,, and Teuling}]{seneviratne_investigating_2010}
Seneviratne, S.~I., T.~Corti, E.~L. Davin, M.~Hirschi, E.~B. Jaeger, I.~Lehner, B.~Orlowsky, and A.~J. Teuling, 2010: Investigating soil moisture–climate interactions in a changing climate: {A} review. \textit{Earth-Science Reviews}, \textbf{99~(3)}, 125--161, \doi{10.1016/j.earscirev.2010.02.004}.

\bibitem[{Seneviratne et~al.(2013)}]{seneviratne_impact_2013}
Seneviratne, S.~I., and Coauthors, 2013: Impact of soil moisture-climate feedbacks on {CMIP5} projections: {First} results from the {GLACE}-{CMIP5} experiment. \textit{Geophysical Research Letters}, \textbf{40~(19)}, 5212--5217, \doi{10.1002/grl.50956}.

\bibitem[{Sherwood et~al.(2020)}]{sherwood_assessment_2020}
Sherwood, S.~C., and Coauthors, 2020: An {Assessment} of {Earth}'s {Climate} {Sensitivity} {Using} {Multiple} {Lines} of {Evidence}. \textit{Reviews of Geophysics}, \textbf{58~(4)}, e2019RG000\,678, \doi{10.1029/2019RG000678}.

\bibitem[{Singh et~al.(2017)Singh, Kuang, Maloney, Hannah,, and Wolding}]{singh2017increasing}
Singh, M.~S., Z.~Kuang, E.~D. Maloney, W.~M. Hannah, and B.~O. Wolding, 2017: Increasing potential for intense tropical and subtropical thunderstorms under global warming. \textit{Proceedings of the National Academy of Sciences}, \textbf{114~(44)}, 11\,657--11\,662.

\bibitem[{Singh and O'Gorman(2013)Singh, and O'Gorman}]{singh_influence_2013}
Singh, M.~S., and P.~A. O'Gorman, 2013: Influence of entrainment on the thermal stratification in simulations of radiative-convective equilibrium. \textit{Geophysical Research Letters}, \textbf{40~(16)}, 4398--4403, \doi{10.1002/grl.50796}.

\bibitem[{Sobel and Bretherton(2000)Sobel, and Bretherton}]{sobel_modeling_2000}
Sobel, A.~H., and C.~S. Bretherton, 2000: Modeling {Tropical} {Precipitation} in a {Single} {Column}. \textit{Journal of Climate}, \textbf{13~(24)}, 4378--4392, \doi{10.1175/1520-0442(2000)013<4378:MTPIAS>2.0.CO;2}.

\bibitem[{Tamarin-Brodsky et~al.(2019)Tamarin-Brodsky, Hodges, Hoskins,, and Shepherd}]{tamarin-brodsky_dynamical_2019}
Tamarin-Brodsky, T., K.~Hodges, B.~J. Hoskins, and T.~G. Shepherd, 2019: A {Dynamical} {Perspective} on {Atmospheric} {Temperature} {Variability} and {Its} {Response} to {Climate} {Change}. \textit{Journal of Climate}, \textbf{32~(6)}, 1707--1724, \doi{10.1175/JCLI-D-18-0462.1}.

\bibitem[{Tamarin-Brodsky et~al.(2020)Tamarin-Brodsky, Hodges, Hoskins,, and Shepherd}]{tamarin-brodsky_changes_2020}
Tamarin-Brodsky, T., K.~Hodges, B.~J. Hoskins, and T.~G. Shepherd, 2020: Changes in {Northern} {Hemisphere} temperature variability shaped by regional warming patterns. \textit{Nature Geoscience}, \textbf{13~(6)}, 414--421, \doi{10.1038/s41561-020-0576-3}.

\bibitem[{Tian et~al.(2023)}]{tian_radiation_2023}
Tian, Y., and Coauthors, 2023: Radiation as the dominant cause of high-temperature extremes on the eastern {Tibetan} {Plateau}. \textit{Environmental Research Letters}, \textbf{18~(7)}, 074\,007, \doi{10.1088/1748-9326/acd805}.

\bibitem[{Vallis(2017)}]{vallis_atmospheric_2017}
Vallis, G.~K., 2017: \textit{Atmospheric and {Oceanic} {Fluid} {Dynamics}: {Fundamentals} and {Large}-{Scale} {Circulation}}. 2nd ed., Cambridge University Press, Cambridge, \doi{10.1017/9781107588417}, \urlprefix\url{https://www.cambridge.org/core/product/41379BDDC4257CBE11143C466F6428A4}.

\bibitem[{Vallis et~al.(2018)}]{vallis_isca_2018}
Vallis, G.~K., and Coauthors, 2018: Isca, v1.0: a framework for the global modelling of the atmospheres of {Earth} and other planets at varying levels of complexity. \textit{Geoscientific Model Development}, \textbf{11~(3)}, 843--859, \doi{10.5194/gmd-11-843-2018}.

\bibitem[{Vasseur et~al.(2014)}]{vasseur_increased_2014}
Vasseur, D.~A., and Coauthors, 2014: Increased temperature variation poses a greater risk to species than climate warming. \textit{Proceedings of the Royal Society B: Biological Sciences}, \textbf{281~(1779)}, 20132\,612, \doi{10.1098/rspb.2013.2612}.

\bibitem[{Vogel et~al.(2019)Vogel, Donat, Alexander, Meinshausen, Ray, Karoly, Meinshausen,, and Frieler}]{vogel_effects_2019}
Vogel, E., M.~G. Donat, L.~V. Alexander, M.~Meinshausen, D.~K. Ray, D.~Karoly, N.~Meinshausen, and K.~Frieler, 2019: The effects of climate extremes on global agricultural yields. \textit{Environmental Research Letters}, \textbf{14~(5)}, 054\,010, \doi{10.1088/1748-9326/ab154b}.

\bibitem[{Zeppetello et~al.(2022)Zeppetello, Battisti,, and Baker}]{zeppetello2022physics}
Zeppetello, L. R.~V., D.~S. Battisti, and M.~B. Baker, 2022: The physics of heat waves: What causes extremely high summertime temperatures? \textit{Journal of Climate}, \textbf{35~(7)}, 2231--2251.

\bibitem[{Zhang and Boos(2023)Zhang, and Boos}]{zhang_upper_2023}
Zhang, Y., and W.~R. Boos, 2023: An upper bound for extreme temperatures over midlatitude land. \textit{Proceedings of the National Academy of Sciences}, \textbf{120~(12)}, e2215278\,120, \doi{10.1073/pnas.2215278120}.

\bibitem[{Zhang and Fueglistaler(2020)Zhang, and Fueglistaler}]{zhang_how_2020}
Zhang, Y., and S.~Fueglistaler, 2020: How {Tropical} {Convection} {Couples} {High} {Moist} {Static} {Energy} {Over} {Land} and {Ocean}. \textit{Geophysical Research Letters}, \textbf{47~(2)}, e2019GL086\,387, \doi{10.1029/2019GL086387}.

\bibitem[{Zhang et~al.(2021)Zhang, Held,, and Fueglistaler}]{zhang_projections_2021}
Zhang, Y., I.~Held, and S.~Fueglistaler, 2021: Projections of tropical heat stress constrained by atmospheric dynamics. \textit{Nature Geoscience}, \textbf{14~(3)}, 133--137, \doi{10.1038/s41561-021-00695-3}.

\end{thebibliography}

\end{document}